\RequirePackage{fix-cm}

\documentclass[twocolumn,epjc3]{svjour3}          % twocolumn

\RequirePackage[T1]{fontenc}

\smartqed  % flush right qed marks, e.g. at end of proof
\RequirePackage{epsfig}
\RequirePackage{subfigure}
\RequirePackage{graphics,color}
\RequirePackage[none]{hyphenat}
\RequirePackage{widetext}
\RequirePackage{amsbsy}
\RequirePackage{mathtools}
\RequirePackage{amsmath}
\RequirePackage{bm}
\RequirePackage{graphicx}
\usepackage{epstopdf}
\RequirePackage{mathptmx}      % use Times fonts if available on your TeX system
\RequirePackage{flushend}
\RequirePackage[numbers,sort&compress]{natbib}
\RequirePackage[colorlinks,citecolor=blue,urlcolor=blue,linkcolor=blue]{hyperref}
\newcommand{\nn}{\nonumber}
\newcommand{\Tr}{{\mathrm{Tr}}}
\def \l{\left(}
\def \r{\right)}

\journalname{Eur. Phys. J. Plus }

\begin{document}

\title{Exploring axial $U(1)$ restoration in a modified 2+1 flavor Polyakov quark meson model}

\subtitle{}

\author{Suraj Kumar Rai\thanksref{e1,addr1}
        \and
        Vivek Kumar Tiwari\thanksref{e2,addr1} %etc.
}

%\thankstext[$\star$]{t1}{Thanks to the title}
\thankstext{e1}{e-mail:surajrai050@gmail.com}
\thankstext{e2}{e-mail:vivekkrt@gmail.com}

\institute{Department of Physics, University of Allahabad, Allahabad 211002, India.\label{addr1}}

%\date{Received: date / Accepted: date}
% The correct dates will be entered by the editor

\maketitle

\begin{abstract}
\sloppy
We report on the $U_{A}(1)$ symmetry restoration resulting due to temperature dependence of the coefficient c(T) for the Kobayashi-Maskawa-'t Hooft determinant (KMT) term in a modified 2+1 flavor Polyakov loop quark meson model having fermionic vacuum correction term (PQMVT).Temperature dependence of KMT coupling c(T) drives the non-strange condensate melting to significantly smaller temperatures in comparison to the constant c case.Further due to c(T), $m_{\eta^{\prime}}$ decreases from its vacuum value by 220 MeV near T=176 MeV after the chiral transition ($T_{c}^{\chi}=154.9$ MeV).This is similar to the $\eta^{\prime}$ in-medium mass drop of at least 200 MeV as reported by Csorgo and Vertesi in Ref \cite{Csorgo,Vertesi}, as an experimental signature of the effective restoration of $U_A(1)$ symmetry. 
The pseudoscalar mixing angle $\theta_p$ achieves anti-ideal mixing in the influence of c(T). The $\eta $ meson becomes light quark system ($ \eta_{NS} $) at T=176 MeV and changes its identity with $ \eta' $ meson which becomes strange quark system ($ \eta_{S} $). The degenerated temperature variations of $\sigma$, $\pi$ meson masses merges with the temperature variations of the masses of degenerated $a_{0}$, $\eta$ mesons near 275 MeV. It means that for c(T) when $m_{\sigma}=400$ MeV, the $U_{A}(1)$ 
restoration  takes place at 1.75 $T_{c}^{\chi}$=275 MeV.
\end{abstract}

\section{Introduction}
\label{intr}
\sloppy
 The normal hadronic matter under the extreme conditions of 
high temperature and/or density, turns into a collective form of matter 
known as the Quark Gluon Plasma (QGP) when the individual hadrons dissolve 
into their quark and gluon constituents \cite{Shuryak,Rafelski,Svetitsky,Ortmanns:96ea,Muller,Rischke:03}.
Relativistic heavy ion collision experiments at RHIC (BNL), LHC (CERN) and 
the future CBM experiments at the FAIR facility (GSI-Darmstadt) aim to create
and study such a collective state of matter. Study of the different aspects of 
this phase transition, is a tough and challenging task because Quantum Chromodynamics (QCD),
the theory of strong interaction, becomes nonperturbative in the low energy limit.

  It is well known that the quantum chromodynamics (QCD) Lagrangian has 
the global $SU_{V}(3) \times SU_{A}(3) \times U_V(1) \times U_A(1)$ symmetry. 
The $U_V(1)$ symmetry gives baryon number 
conservation. The QCD vacuum reveals itself through the process of spontaneous breakdown of 
chiral symmetry $ SU_{A}(3)$ for the massless quarks in the low
energy limit of QCD  . Chiral condensate works as an order 
parameter and one gets eight massless pseudoscalar mesons as Goldstone bosons for the low
energy hadronic vacuum of the QCD. For small quark masses in the QCD, the  mesons become 
pseudo-Goldstone bosons with small non zero masses in comparison to other hadrons . 
The chiral condensate becomes zero at higher temperature  $T_{c}$ in the chiral limit of zero quark mass and this
leads to chiral symmetry restoring phase transition. For 2+1 flavor of light quark masses, the 
phase transition turns into a smooth crossover. Lattice QCD simulations 
show that the pseudo-critical temperature for such chiral crossover is $T_{c}=154 \pm 9$ 
\cite{LQCDWB2,HotLQCDL}.

QCD also has $U_A(1)$ axial symmetry at the classical level which is  broken explicitly
by the quantum anomaly as shown by 't Hooft \cite{tHooft:76prl}. In QCD vacuum, Instanton 
effects explicitly break the $U_A(1)$ to $Z_{A}(3)$. $U_A(1)$ symmetry breaking can be understood
in terms of Kobayashi-Maskawa-'t Hooft (KMT) determinant \cite{tHooft:76prl,Kobaya} describing the flavor 
mixing six quark interaction that gets contributions from fluctuations in the topological charge.
Flavor mixing removes the degeneracy among several mesons.The pseudoscalar singlet $\eta'$ meson
acquires a mass of about 1 GeV. In the absence of $U_A(1)$ anomaly, $\eta'$  would have been degenerate
with $\pi$ in $U(3)$ as a ninth Goldstone boson in the spontaneously broken chirally symmetric phase. 
Since the fluctuations in the topological charge are suppressed at temperatures far above the
chiral crossover temperature $T_{c}$ \cite{Gross,Yaffe,TSchaf},it is expected that the axial anomaly vanishes and  
$U_A(1)$ symmetry gets effectively restored for temperature $T_{U(1)}>>T_{c}$.Shuryak \cite{Shuryak1} has discussed 
two scenarios : the complete chiral symmetry  $ SU_{A}(3) \times U_A(1)$ is restored in scenario one
for $T_{U(1)}>>T_{c}$ well inside the quark gluon plasma region and in scenario two $T_{U(1)} \sim T_{c}$. 
Further, it has been argued that mass splitting between 
pion and pseudoscalar singlet meson could become small immediately above the chiral crossover \cite{Jkapusta}.
In Ref \cite{Csorgo,Vertesi}, it is reported that the in-medium mass of pseudoscalar singlet meson drops by 
at least 200 MeV which supposedly is an experimental signature of effective restoration of  $U_A(1)$ symmetry . 

Some lattice QCD calculations have addressed the issue of effective restoration of $U_A(1)$ 
symmetry. Recently $U_A(1)$ restoration was explored by computing the degeneration of
pion and $a_{0}$-meson screening masses using improved $ p^{4}$ staggered fermions \cite{ChengLQCD}. 
In a two-flavor simulation with overlap and domain wall fermions\cite{CossuL,CossuLQCD}, the correlators of pion
and pseudo-scalar singlet meson have been observed to be degenerate in the chiral symmetric phase .
For the 2+1 domain wall fermions, Ref \cite{TBHATTA,BazLQCD} find effective restoration of $U_A(1)$ at a larger
temperature of 196 MeV while Refs \cite{Sharma, Dick} do not see restoration  even at 1.5 times the crossover 
temperature using highly improved staggered fermions. The interplay of $U_A(1)$ axial symmetry restoration and 
chiral crossover transition has been explored in several effective model settings like Dyson-Schwinger approach 
using quark-gluon interaction \cite{Smekal,Alfok,Benic,SBenic}, in the Nambu-Jona-Lasinio (NJL) model
\cite{Kunihiro,Costa:04,Costa:05,Costa:09,Chen,Zhang,Powell,Ruivo, Bratovic}
 and linear sigma models\cite{Bielich:00prl,Rischke:00,Fejos,Eser} which 
become quark-meson (QM) models \cite{Schaefer:09,Schaefer:08ax, Mitter,Herbst} when coupled with quarks , 
in the Polyakov loop augmented Nambu-Jona-Lasinio (PNJL) models and Polyakov loop enhanced 
quark-meson (PQM) models. PNJL and PQM models combine  the features of spontaneous breakdown of both 
chiral symmetry as well as the center $Z(3)$ symmetry of QCD in one single 
model (see for example \cite{Costa:09,Ruivo,Schaefer:07,gupta,Schaefer:09ax,Schaefer:09wspax,H.mao09,
Herbst:2010rf,Pawl:Schaef,Marko:2010cd,kahara,Digal:01,Pisarski:00prd,Dumitru,Ratti:06,
Ratti:07,Haas,Aichelin,Kovacs,  Ghosh,Hansen:07,Ratti:07npa,Tamal:06,Sasaki:07,Hell:08,Abuki:08,Ciminale:07,
Fu:07,Kfuku:04plb,Fukushima:08d77,Fukushima:08d78,Fukushima:09,Contrera,
nonlocal,Odilon,Contrera:NLM,Hiller}). In these models chiral condensate and Polyakov loop are simultaneously 
coupled to the quark degrees of freedom.
In the high temperature/density regime, QCD vacuum shows color confinement-deconfinement phase transition 
for the infinitely heavy quarks where the center symmetry $Z(3)$ of color gauge group $SU_c(3)$ gets 
spontaneously broken and the expectation value of the Wilson line (Polyakov loop) serves as the order
parameter \cite{Polyakov:78plb} as it is related to the free energy of a static color charge .
Since the center symmetry is always explicitly broken in the presence of dynamical quarks in the system, 
one regards the Polyakov loop as an approximate order parameter for the confinement-deconfinement transition
\cite{Pisarski:00prd}.

The effect of confinement physics i.e that of the Polyakov loop 
potential on the relation of $SU(3)$ chiral symmetry and $U_A(1)$ axial symmetry above chiral crossover temperature
has been explored in the 2+1 flavor PQM model in Ref.\cite{gupta}.Earlier,in the 2+1 flavor QM model, 
Schaefer et. al.\cite{Schaefer:09,Schaefer:08ax} had studied the interplay of these symmetries at higher temperatures.
When renormalized fermionic vacuum fluctuation is included in their thermodynamic potential, the two flavor 
QM model and PQM model \cite{Skokov:2010sf, Vivek:12,VKTR}, become effective QCD-like because 
now models can reproduce the second order chiral phase transition at $\mu=0$ as expected from the universality 
arguments \cite{Wilczek} for the two massless flavors of QCD. In earlier works of implementing fermionic vacuum correction in 
the 2+1 flavor PQM model, the renormalization scale independence of the model parameters and meson masses used to be implemented 
numerically \cite{Schaef:12, Sandeep}. Recently, explicit renormalization scale independent 
analytical expressions for the  model parameters, meson masses and mixing angles have been found for the 2+1 flavour
PQM model in the Ref.  \cite{Trvk} where the effect of fermionic vacuum correction, on the interplay of $SU_A(3)$ chiral symmetry
and $U_A(1)$ symmetry restoration has been explored. We will be considering a modification of this model in the present work.

It is very relevant to ask if the $U_A(1)$ symmetry,which is broken in the vacuum by instantons at the 
quantum level, is still broken in the chiral symmetric phase \cite{Meggiolaro,fukushima:01}.If the $U_A(1)$ symmetry 
breaking strength decreases with temperature, it is pertinent to ask what fraction of it remains above the critical temperature 
and whether and when this symmetry is restored \cite{Ruivo}. It has been pointed out that there are phenomenological consequences
for the nature of phase transition depending on the degree of anomaly present at the critical temperature \cite{BazLQCD}. 
Temperature dependence in the form of decreasing exponential for the axial anomaly term was originally proposed by Kunihiro in Ref. 
\cite{Kunihiro} and later its phenomenological effects for $U_A(1)$ restoration was explored by Costa et. al. in Ref.
\cite{Costa:04,Costa:05} for the NJL model. Another form of exponentially decreasing
temperature dependence of anomaly coefficient and its consequences for the $U_A(1)$ symmetry restoration  has been explored in 
Ref. \cite{Ruivo} in the context of SU(2) EPNJL model. 

Here, we point 
out that the evolution of $U_A(1)$ anomaly and its relation to the chiral symmetry in the extended linear sigma model, QM model
has also been investigated in the powerful framework of nonperturbative functional renormalization group (FRG) method 
\cite{Fejos,Eser,Mitter,Yin,Kamikado,Jiang,Fejos:16,Heller,Rennecke,Fejos:18} where the 
bosonic and fermionic quantum and thermal fluctuations are included in the FRG effective action through running the RG 
scale from the ultraviolet scale to the infrared limit, which, as an advantage, can automatically guarantee 
the Nambu-Goldstone theorem in the symmetry breaking phase. The thermal evolution of $U_A(1)$ factor when explored by the chiral 
invariant expansion technique \cite{Fejos}, shows that mesonic fluctuations not only produce condensate dependent anomaly
coefficient, but depending on region of the parameter space, they can either strengthen or weaken the KMT determinant term 
as the temperature increases. In Ref. \cite{Fejos:16,Fejos:18}, it was found that mesonic fluctuations strengthen the anomaly
towards $T_{c}$. While recent lattice simulations, show that the anomaly does not restore until around  $1.5 \ T_{c}$ \cite{Sharma,Dick}.
In this light, it becomes an important question whether temperature dependence of KMT coupling that arises from the instanton 
effects, can offset the effect of the mesonic fluctuations.

In the present work, we consider a temperature dependence of the
coefficient c(T) for the Kobayashi-Maskawa-'t Hooft (KMT) determinant term in the 2+1 flavor PQM model with 
fermionic vacuum correction (PQMVT). Phenomenological form of strong
temperature suppression for the KMT coupling strength was inferred from the lattice data on  ($\pi, a_{0}$) screening masses\cite{ChengLQCD}.
Its effect on the $U_{A}(1)$ symmetry restoration has already been studied in the 2+1 flavor 
entanglement-PNJL (EPNJL) model \cite{Ishii,Ishiky} where they find degeneration in the masses of $a_0$ and $\pi $ mesons
at T=180 MeV=$1.1 \ T_{c}$ with $T_{c}=163$ MeV which means $U_A(1)$ restoration takes places immediately after chiral crossover.
We do not have such investigations in the PQM model for meson masses calculated from the curvature of effective potential. It is worthwhile
to have such a detailed investigation in the PQMVT model also for obtaining qualitative and quantitative trends of $U_A(1)$ 
symmetry restoration that may facilitate comparison with other studies. Further it is to be noted  that the mesons 
in the PQM model are the degrees of freedom included in the Lagrangian form the outset while those are generated by some 
prescription in the NJL model \cite{Costa:05} and the $\eta'$ is not a well defined quantity \cite{fukushima:01} which 
becomes unbound for higher temperatures.

%We will be computing the modification in the $U_{A}(1)$ symmetry restoration pattern when the constant coupling
% becomes the temperature dependent coupling strength C(T) in the MPQM model. \langle%C
% We shall compare how the temperature
%dependence of C influences the interplay of chiral symmetry restoration and the $U_A(1)$ restoration trends.
%We will also study how the temperature dependence of the coupling strength C(T) leads to the modification 
%in the role of the $U_A(1)$ anomaly in determining the isoscalar masses and mixing angles for the pseudoscalar 
%($\eta$ and $\eta'$) and scalar ($\sigma$ and $f_0$) meson complex. 

This paper has the following arrangement . Sec.\ref{sec:model}, explains the formulation of the model.
In subsection \ref{subsec:KMT}, we have explained the introduction of temperature dependence for KMT coupling.
In subsection \ref{subsec:Plgtp}, the used form of Polyakov loop potential has been explained. The   
Sec.\ref{sec:potmf} describes the mean field grand potential with fermionic vacuum correction.
The Sec.\ref{sec:mixang} gives the model formulae of meson masses and mixing angles in the vacuum and the
finite temperature/density medium. The Sec. \ref{sec:parameter} gives the details of parameter fixing.
In Sec.\ref{sec:eploop}, we will be discussing the numerical results and plots for temperature variations 
of meson masses and mixing angles. Summary and conclusion is presented in the last Sec.\ref{sec:smry}.
%The fermionic vacuum correction also generates, modifications in the vacuum masses and mixing angles of the
%scalar and pseudo-scalar mesons. 
 %%%%%%%%%%%%%%%%%%%%%%%%%%%%%%%%%%%%%%%%%%%%%%%%%%%%%%%%%%%%%%%%%%%
%%%   Model Formulation
%%% 
%%%%%%%%%%%%%%%%%%%%%%%%%%%%%%%%%%%%%%%%%%%%%%%%%%%%%%%%%%%%%%%%%%%
\section{Model Formulation}
\label{sec:model} 
 
In the model \cite{gupta,Schaefer:09wspax,H.mao09,Schaefer:09ax} , three flavor of quarks are coupled to the 
$SU_V(3) \times SU_A(3)$ symmetric mesonic fields together with 
temporal component of gauge field represented by the Polyakov loop 
potential. Thermal expectation value of color trace of Wilson loop in temporal 
direction defines the Polyakov loop field $\Phi$ as
\begin{equation}
\Phi(\vec{x}) = \frac{1}{N_c} \langle \Tr_c L(\vec{x})\rangle, \qquad \qquad  \bar\Phi(\vec{x}) =
\frac{1}{N_c} \langle \Tr_c L^{\dagger}(\vec{x}) \rangle
\end{equation}
where $L(\vec{x})$ is a matrix in the fundamental representation of the 
$SU_c(3)$ color gauge group.
\begin{equation}
\label{eq:Ploop}
L(\vec{x})=\mathcal{P}\mathrm{exp}\left[i\int_0^{\beta}d \tau
A_0(\vec{x},\tau)\right]
\end{equation}
Here $\mathcal{P}$ is path ordering,  $A_0$ is the temporal component of vector 
field and $\beta = T^{-1}$ \cite{Polyakov:78plb}. In accordance to
Ref. \cite{Ratti:06,Ratti:07} we consider an homogeneous Polyakov loop field $\Phi(\vec{x})=\Phi$=constant and
$\bar\Phi(\vec{x})=\bar\Phi$=constant.
 
The model Lagrangian has quarks, mesons, couplings 
and Polyakov loop potential ${\cal U} \left( \Phi, \bar\Phi, T \right)$ as :

\begin{equation}
\label{eq:Lag}
{\cal L}_{PQM} = {\cal L}_{QM} - {\cal U} \big( \Phi , \bar\Phi , T \big) 
\end{equation}
where the Lagrangian in Quark Meson Chiral Sigma model
\begin{eqnarray}
\label{eq:Lqms}
{\cal L}_{QM} =  \bar{q_f} \big( i \gamma^{\mu}D_{\mu}  - g\; T_a\big( \sigma_a 
+ i\gamma_5 \pi_a\big) \big) q_f  + {\cal L}_{m} 
\end{eqnarray}
Quarks couple with the uniform temporal background gauge 
field as the following 
$D_{\mu} = \partial_{\mu} -i A_{\mu}$ 
and  $A_{\mu} = \delta_{\mu 0} A_0$ (Polyakov gauge), where 
$A_{\mu} = g_s A^{a}_{\mu} \lambda^{a}/2$ with vector potential $A^{a}_{\mu}$ for color gauge field.
$g_s$ is the $SU_c(3)$ gauge coupling. $\lambda_a$ are Gell-Mann matrices in the color 
space, $a$ runs from $1 \cdots 8$. $q_f=(u,d,s)^T$ denotes the quarks 
of three flavors and three colors. $T_a$ represent 9 generators of $U(3)$ flavor symmetry with 
$T_a = \frac{\lambda_a}{2}$ and $a=0,1 \dots ~8$, here $\lambda_a$ are 
standard Gell-Mann matrices in flavor space with $\lambda_0=\sqrt{\frac{2}{3}}\ \bf 1$. $g$ is the flavor blind 
Yukawa coupling that couples the three flavor of quarks with nine 
mesons in the scalar ($\sigma_a, J^{P}=0^{+}$) and pseudo scalar 
($\pi_a, J^{P}=0^{-}$) sectors. 
The quarks acquire mass though  spontaneous breakdown of chiral symmetry .  

The mesonic part of the Lagrangian has the following form
\begin{eqnarray}  
\label{eq:Lagmes}
{\cal L}_{m} & = &\Tr \left( \partial_\mu M^\dagger \partial^\mu
    M \right)
  - m^2 \Tr ( M^\dagger M) -\lambda_1 \left[\Tr (M^\dagger M)\right]^2 \nn \\
  &&  - \lambda_2 \Tr\left(M^\dagger M\right)^2
  +c   \big[\text{det}(M) +\text{det} (M^\dagger) \big] \nn \\
  && + \Tr\left[H(M + M^\dagger)\right].
\end{eqnarray}
The chiral field $M$ is a $3 \times 3$ complex matrix comprising of 
the nine scalars $\sigma_a$ and the nine pseudo scalar $\pi_a$ mesons.
\begin{equation}
\label{eq:Mfld}
M = T_a \xi_a= T_a(\sigma_a +i\pi_a)
\end{equation} 
The generators follow $U(3)$ algebra $\left[T_a, T_b\right]  = if_{abc}T_c$ 
and $\left\lbrace T_a, T_b\right\rbrace  = d_{abc}T_c$ where 
$f_{abc}$ and $d_{abc}$ are standard antisymmetric and symmetric 
structure constants respectively with 
$f_{ab0}=0$ and $d_{ab0}=\sqrt{\frac{2}{3}} \ {\bf 1}\ \delta_{ab}$ 
and matrices are normalized as $\Tr(T_a T_b)=\frac{\delta_{ab}}{2}$. c is the
KMT coupling, a parameter considered in general as constant.

The $SU_L(3) \times SU_R(3)$ chiral symmetry is explicitly broken by 
the explicit symmetry breaking term 
\begin{equation}
H = T_a h_a
\end{equation}
Here $H$ is a $3 \times 3$ matrix with nine external parameters. 
The $\xi$ field which denotes both the scalar as well as pseudo scalar mesons,
picks up the nonzero vacuum expectation value, $\bar{\xi}$ for the scalar mesons 
due to the spontaneous breakdown of the chiral symmetry while the pseudo scalar mesons
have zero vacuum expectation value. Since $\bar{\xi}$ must have the quantum numbers of the vacuum, 
explicit breakdown of the chiral symmetry is only possible with 
three nonzero parameters $h_0$, $h_3$ and $h_8$. We are neglecting 
isospin symmetry breaking hence we choose $h_0$, $h_8 \neq 0$.
This leads to the $2+1$ flavor symmetry breaking scenario with 
nonzero condensates $\bar{\sigma_0}$ and $\bar{\sigma_8}$. 

\subsection{Temperature dependence of KMT Coupling }
%and Mean field grand potential}
%%%%%%%%%%%%%%%%%%%%%%%%%%%%%%%%%%%%%%%%%%%%%%%%%%%%%%%%%%%%%%%%%%%%%%%%%%%%%%
\label{subsec:KMT}

The density of the instantons gets screened by the medium at
finite temperature $T$ . Since the coupling $c$ of the KMT
interaction is proportional to the instanton density , it 
gets reduced as $T$ increases. Such suppression $S(T)\equiv c(T)/c(0) $ was discussed by Pisarski 
and Yaffe for high temperatures \cite{Yaffe}, say $T \geq 2 T_{c} $ for the chiral crossover transition pseudocritical
temperature $T_{c}$, by calculating the Debye-type screening 
\begin{equation}
\label{eq:instasup}
S(T)=exp[-\pi^{2} \rho^{2} T^{2}(\frac{2}{3}N_{c}+\frac{1}{3}N_{f})] \\
=exp[-\frac{T^2}{b^2}]
\end{equation}
Here $N_{f}(N_{c})$ is the number of flavors (colors) and the typical instanton radius $\rho$ is about $\frac{1}{3}$
fm, hence the suppression parameter b is about $0.70 T_{c}$ for $N_{f}=N_{c}=3$. The temperature dependence
of the instanton density was theoretically estimated by the instanton-liquid model \cite{Shuryak1} after Pisarski-Yaffe discussion on $S(T)$
but this estimation is applicable only for $ T \geq 2 T_{c} $ . Due to this reason, Ruivo et. al. in Ref. \cite{Ruivo} used a phenomenological
Wood-Saxon form $1/(1+exp((T-T_{1}^\prime)/b_{1}^\prime)) $ with two parameters ${T_{1}^\prime}$ and  ${b_{1}^\prime}$
for the instanton suppression $S(T)$ factor. Further, lattice QCD study suggests that at $T\sim1.5 \ T_{c}$, the origin 
of  $U_A(1)$ breaking results due to the dilute gas of weakly interacting instantons and 
anti-instantons \cite{Sharma,Dick}. It would be interesting to explore, how the suppression of instanton density and    
the diminishing of the KMT coupling in the vicinity of the chiral crossover transition temperature $T_{c}$ 
(say $T \sim 0.7 \ T_{c} $), influences the restoration of chiral symmetry and its interplay with the $U_A(1)$
symmetry restoration in the temperature range $T \sim (0.7 \ T_{c} \to 2.0 \ T_{c})  $ .

In our model, we  consider  the following temperature dependence for the coefficient  c(T) :    

\begin{eqnarray}
\label{eq:CoupT}
c(T)= 
\begin{dcases}
c(0),                       & (T < T_{1})\\ 
c(0) \ exp[{-\frac{(T-T_{1})^{2}}{b_{1}^{2}}}], & (T > T_{1}) 
\end{dcases}
\end{eqnarray}

The above parameterization of KMT coupling strength was introduced very recently for exploring the 
$U_A(1)$ symmetry restoration in the context of the entanglement-PNJL (EPNJL) model \cite{Ishii,Ishiky} where
the parameters $ T_{1} $ and $b_{1}$ are determined from  LQCD data on pion and $a_{0}$-meson
screening masses . The $T$ dependence of Eq.~(\ref{eq:CoupT}) is consistent with the Pisarski and Yaffe suppression of the
instanton density for high $T \ge 2 T_{c}$ and is similar to the Woods-Saxon form \cite{Yaffe,Gross}. In our model
setting, we shall determine the parameters $ T_{1} $ and $b_{1}$ by fitting the subtracted chiral order parameter
obtained in our calculation to the Wuppertal group LQCD data \cite{LQCDWB2} for the same . 
We have considered temperature dependence of subtracted chiral order parameter for fitting it with the LQCD data
because other quantities depend on it and it is simple to understand and numerically implement in our model calculation.
Further the physics of temperature dependent reduction in axial anomaly is interlinked with the chiral symmetry restoration
and confinement-deconfinement transition. 
%Since it is broken by the quantum effects, the $U_A(1)$ axial 
%which otherwise is a symmetry of the classical Lagrangian, becomes
%anomalous~\cite{Weinberg:75} and gives large mass to $\eta'$ meson 
%($m_{\eta'} = 940$ MeV).

% In the absence of $U_A(1)$ anomaly, $\eta'$ 
%meson would have been the ninth pseudo scalar Goldstone boson, 
%resulting due to the spontaneous break down of the chiral $U_A(3)$ 
%symmetry. The entire pseudo scalar nonet corresponding to the spontaneously 
%broken $U_A(3)$, would consist of the three $\pi$, four $K$, $\eta$ 
%and $\eta'$ mesons, which are the massless pure Goldstone
%modes when $H = 0$ and they become pseudo Goldstone modes after 
%acquiring finite mass due to nonzero $H$ in different symmetry 
%breaking scenarios. The particles coming from octet 
%($a_0$, $f_0$, $\kappa$) and singlet ($\sigma$) representations 
%of $SU_V(3)$ group, constitute scalar nonet 
%($\sigma$, $a_0$, $f_0$, $\kappa$). In order to study the chiral 
%symmetry restoration at high temperatures, we will be investigating 
%the trend of convergence in the masses of chiral partners occurring 
%in pseudo scalar ($\pi$, $\eta$, $\eta'$, $K$) and scalar 
%($\sigma$, $a_0$, $f_0$, $\kappa$) nonets, in the $2+1$ flavor 
%symmetry breaking scenario.

%%%%%%%%%%%%%%%%%%%%%%%%%%%%%%%%%%%%%%%%%%%%%%%%%%%%%%%%%%%%%%%%%%%%%%%%%%%%%%
%%%%%%%%%%%%%%%%%%%%%%%%%%%%%%%%%%%%%%%%%%%%%%%%%%%%%%%%%%%%%%%%%%%%%%%%%%%%%%
\subsection{Polyakov Loop Potential }
%and Mean field grand potential}
%%%%%%%%%%%%%%%%%%%%%%%%%%%%%%%%%%%%%%%%%%%%%%%%%%%%%%%%%%%%%%%%%%%%%%%%%%%%%%
\label{subsec:Plgtp}
The effective Polyakov loop potential ${\cal U} \left( \Phi, \bar\Phi, T \right)$ 
is constructed such that it reproduces thermodynamics of pure glue 
theory on the lattice for temperatures upto about twice the 
deconfinement phase transition temperature. The functional form of Polyakov loop potential
has various possibilities. The simplest 
Polyakov loop effective potential as introduced in \cite{Pisarski:00prd,Dumitru,Ratti:06} has a polynomial form.
The polynomial form of Polyakov loop potential in Ref. \cite{Ratti:06}
produces some unwanted properties such as negative Susceptibilities \cite{Sasaki:07}. We are
using logarithmic parameterization of Polyakov loop potential which comes from the SU(3) 
Haar measure of the group integration \cite{Ratti:07} and the negative Susceptibilities problem is 
also absent in this case . The logarithmic Polyakov loop potential is the following
 
\begin{eqnarray}
\label{eq:logpot}
\frac{{\cal U_{\text{log}}}\left(\Phi,\bar\Phi, T \right)}{T^4} &=&-\frac{1}{2} \ a(T) \Phi \bar\Phi  +
b(T) \, \mbox{ln}[1-6\bar\Phi \Phi \nn \\&&+4(\bar\Phi^{3}+ \Phi^3)-3(\bar\Phi \Phi)^2]
\end{eqnarray}

where the parameters are the following

\begin{equation} 
  a(T) =  a_0 + a_1 \left(\frac{T_0}{T}\right) + a_2 \left(\frac{T_0}{T}\right)^2 , \ b(T)=
b_3 \left(\frac{T_0}{T}\right)^3 .
\end{equation}

where constants have the values 
\begin{eqnarray*}
&& a_0 = 3.51,\ a_1= -2.47, \ a_2= -2.47, \ \text{and} \  b_3= -1.75\ 
\end{eqnarray*}

%\begin{widetext}
For pure gauge Yang Mills theory, the deconfinement phase transition is first order and the critical temperature is 
$T_0=270$ MeV. In the presence of dynamical quarks, the first order transition turns to
a crossover and  $T_0$ gets reduced depending on number of quark flavours and chemical potential 
\cite{Schaefer:07,Herbst:2010rf,Pawl:Schaef, Haas}. As given  in Ref. \cite{Schaefer:07}, at zero
chemical potential $T_{0}=187$ for the 2+1 quark flavors. We should note that the change of $T_0$ does not modify the
functional form of the Polyakov loop effective potential. Due to this, the expectation value of the Polyakov loop potential 
fails to coincide with the minimum of the effective potential because $\Phi(T)$ is found by the minimization of the
grand-canonical potential (which includes the quark-antiquark contribution also) and not of only $\cal U_{\text{YM}}$ 
One has to look for a possible effect of quarks on the Polyakov loop effective potential out of its minimum \cite{Aichelin}.
  
  In order to find the gluonic contribution to the pressure $P_{\text{gluon}}$ =
$-\cal{U_{\text{YM}}}$ $( T; \Phi(T) \bar{\Phi}(T)) $, one should introduce quark back reaction onto $\cal {U}$, not 
limited to their effect at the minimum of the potential. 
In the context of the functional renormalization group (FRG) equations applied to QCD, Ref. \cite{Haas}
has compared the pure gauge effective potential $\cal U_{\text{YM}}$ to the ``glue'' potential $\cal U_{\text{glue}}$ where
quark polarization has been included in the gluon propagator. They have found significant differences between the two 
effective potentials. However, the study observed that the two potentials have the same shape and that they can be mapped
into each other by relating the temperatures of the two systems , $T_{\text{YM}}$ and  $T_{\text{glue}}$ respectively. 
Denoting by $\cal U_{\text{YM}}$ the potential in Eq.~(\ref{eq:logpot}), the improved Polyakov loop potential was constructed
as \cite{Haas}
\begin{eqnarray}
\label{pot:improved}
\frac{{\cal U_{\text{glue}}}\left(\Phi,\bar\Phi, T_{\text{glue}} \right)}{T_{\text{glue}}^4} &=&\frac{{\cal U_{\text{YM}}}
\left(\Phi,\bar\Phi, T_{\text{YM}} \right)}{T_{\text{YM}}^4}
\end{eqnarray}
where the relation between temperature $T_{\text{glue}}$ and  $T_{\text{YM}}$ is given by 
\begin{eqnarray}
\label{rescaled}
\frac{T_{\text{YM}}-T_{\text{YM}}^{c}}{T_{\text{YM}}^{c}}=0.57 \ \frac{T_{\text{glue}}-T_{\text{glue}}^{c}}{T_{\text{glue}}^{c}}
\end{eqnarray}
Here, the critical temperature $T_{\text{YM}}^{c}=270 $ MeV and $T_{\text{glue}}^{c} $ is the transition temperature
for the unquenched case. The coefficient 0.57 comes from the comparison of the two effective potentials. 
$T_{\text{glue}}^{c} $ lies within a range $T_{\text{glue}}^{c} \in [180,270]$. In practice , we use in the right-hand 
side of Eq.~(\ref{eq:logpot}) where $T_{0}$ means $T_{\text{YM}}^{c}$, the replacement $ T \rightarrow T_{\text{YM}}^{c} 
(1+0.57(\frac{T}{T_{\text{glue}}^{c}}-1))$ where $(T \equiv T_{\text{YM}})$ on the left side of arrow and $(T \equiv T_{\text{glue}})$
on the right side . In our calculation, we shall use several values of $T_{\text{glue}}^{c}$ in the range to fit the lattice
data for different parameters depending on different values of $\sigma$ meson mass.

%%%%%%%%%%%%%%%%%%%%%%%%%%%%%%%%%%%%%%%%%%%%%%%%%%%%%%%%%%%%%%%%%%%%%%%%%%%%%%
%%%%%%%   Grand Potential in the Mean Field Approach
%%%%%%%
%%%%%%%%%%%%%%%%%%%%%%%%%%%%%%%%%%%%%%%%%%%%%%%%%%%%%%%%%%%%%%%%%%%%%%%%%%%%%%
\section{Grand Potential in the Mean Field Approach}
\label{sec:potmf}
We are considering a spatially uniform system in thermal equilibrium at finite temperature
$T$ and quark chemical potential $\mu_f \ (f=u,\  d \ \text{and} \ s)$. The partition 
function is written as the path integral over quark/antiquark and meson 
fields \cite{gupta,Schaefer:09}
\begin{eqnarray}
\label{eq:partf}
\mathcal{Z}&=& \mathrm{Tr\, exp}[-\beta (\hat{\mathcal{H}}-\sum_{f=u,d,s} 
\mu_f \hat{\mathcal{N}}_f)] \nn \\
&=& \int\prod_a \mathcal{D} \sigma_a \mathcal{D} \pi_a \int
\mathcal{D}q \mathcal{D} \bar{q} \; \mathrm{exp} \bigg[- \int_0^{\beta}d\tau\int_Vd^3x  \nn \\
&& \bigg(\mathcal{L_{QM}^{E}} 
 + \sum_{f=u,d,s} \mu_{f} \bar{q}_{f} \gamma^0 q_{f} \bigg) \bigg]. 
\end{eqnarray}
where $V$ is the three dimensional volume of the system,
$\beta= \frac{1}{T}$ and the superscript $\mathcal{E}$ denotes the euclidean 
Lagrangian. For three quark flavors, in general, the 
three quark chemical potentials are different. In this work, we 
assume that $SU_V(2)$ symmetry is preserved and neglect the small 
difference in masses of $u$ and $d$ quarks. Thus the quark chemical 
potential for the $u$ and $d$ quarks become equal $\mu_x = \mu_u = \mu_d$.
The strange quark chemical potential is $\mu_y = \mu_s$.
%\cite{CJT,Rischke:00}\cite{Hatsuda:98}\cite{Herpay:05,Herpay:06,Herpay:07}.

	Here, the partition function is evaluated in the mean-field 
approximation \cite{Schaefer:08ax,gupta,Schaefer:09}. 
We replace meson fields by their expectation values 
$\langle M \rangle =  T_0 \bar{\sigma_0} + T_8 \bar{\sigma_8}$ 
and neglect both thermal as well as quantum fluctuations of meson 
fields while quarks and anti quarks are retained as quantum fields. 
Now following the standard procedure as given in 
Refs.~\cite{Kapusta_Gale,Schaefer:07,Ratti:06,Kfuku:04plb},
one can obtain the expression of grand potential as the sum of pure 
gauge field contribution ${\cal U} \left(\Phi, \bar\Phi, T \right)$, 
meson contribution and quark/antiquark contribution evaluated in 
the presence of Polyakov loop,
\begin{eqnarray}
\label{eq:grandp}
\Omega_{\rm MF}(T,\mu_{f})=-\frac{T\ln Z}{V} &= &U(\sigma_x,\sigma_y)+{\cal U_{\text{glue}}} \left(\Phi,\bar\Phi,T \right) \nn \\
&&+ \Omega_{\bar{q}q} (T, \mu_{f})
\end{eqnarray}
The mesonic potential $U(\sigma_x,\sigma_y)$ is obtained from the $U(\sigma_0,\sigma_8)$ after 
transforming the original singlet-octet (0, 8) basis of condensates to
the non strange-strange basis $(x, y)$ as in Refs. \cite{Rischke:00,Schaefer:09,gupta,Schaef:12}.
We write the mesonic potential as
\begin{eqnarray}
\label{eq:mesop}
&& U(\sigma_{x},\sigma_{y}) =\frac{m^{2}}{2}\left(\sigma_{x}^{2} +
  \sigma_{y}^{2}\right) -h_{x} \sigma_{x} -h_{y} \sigma_{y}
 - \frac{c}{2 \sqrt{2}} \sigma_{x}^2 \sigma_{y} \nn \\
 && + \frac{\lambda_{1}}{2} \sigma_{x}^{2} \sigma_{y}^{2}+
  \frac{1}{8}\left(2 \lambda_{1} +
    \lambda_{2}\right)\sigma_{x}^{4} +\frac{1}{8}\left(2 \lambda_{1} +
    2\lambda_{2}\right) \sigma_{y}^{4}\ 
\end{eqnarray}
where
\begin{eqnarray}
\sigma_x &=&
\sqrt{\frac{2}{3}}\bar{\sigma}_0 +\frac{1}{\sqrt{3}}\bar{\sigma}_8, \\
\sigma_y &=&
\frac{1}{\sqrt{3}}\bar{\sigma}_0-\sqrt{\frac{2}{3}}\bar{\sigma}_8.
\end{eqnarray}
The chiral symmetry breaking external fields 
($h_x$, $h_y$) are written in terms of ($h_0$, $h_8$) analogously. In our model when
$c$ becomes $ c(T)$, $U(\sigma_{x},\sigma_{y})$ becomes $U(\sigma_{x},\sigma_{y},T)$.

Further the non strange and strange quark/antiquark decouple and the quark masses are 
\begin{equation}
m_x = g \frac{\sigma_x}{2}, \qquad m_y = g \frac{\sigma_y}{\sqrt{2}}
\end{equation} 
Quarks become massive in symmetry broken phase because of non 
zero vacuum expectation values of the condensates.   
The  quark/antiquark contribution \cite{Skokov:2010sf,Trvk}, is written as 
%\begin{equation}
\begin{eqnarray}
&&\Omega_{\bar{q}q}(T,\mu_{f})=\Omega_{q\bar{q}}^{\rm vac}+\Omega_{q\bar{q}}^{\rm T}=-2\sum_{f=u,d,s} \nn \\
&&\int\frac{d^3 p}{(2\pi)^3}\Big[{N_c E_f} \theta( \Lambda^2-\vec{p}^{\,2})+T \{ \ln g_{f}^{+} + \ln g_{f}^{-}\}\Big]
\label{Omeg_q}
\end{eqnarray}
%\end{equation} 
The first term of the Eq.~(\ref{Omeg_q}) represents the fermion vacuum one loop
contribution, regularized by the ultraviolet cutoff $\Lambda$. The expressions $g_{f}^{+}$ and 
$g_{f}^{-}$ in the presence of Polyakov loop potential are defined  in the second term after 
taking trace over the color space
\begin{eqnarray}
\label{eq:gpls} 
g_{f}^{+} = \Big[ 1 + 3\Phi e^{ -E_{f}^{+} /T} +3 \bar\Phi e^{-2 E_{f}^{+}/T} +e^{-3 E_{f}^{+} /T}\Big]\,
\end{eqnarray}
\begin{eqnarray}
\label{eq:gmns} 
g_{f}^{-} = \Big[ 1 + 3\bar\Phi e^{ -E_{f}^{-} /T} +3 \Phi e^{-2 E_{f}^{-}/T} +e^{-3 E_{f}^{-} /T}\Big] \,
\end{eqnarray}
E$_{f}^{\pm} =E_f \mp \mu_{f} $ and $E_f$ is the
flavor dependent single particle energy of quark/antiquark and $m{_f}$ is the mass of the given quark flavor.
\begin{equation}
E_f = \sqrt{p^2 + m{_f}{^2}}
\end{equation} 

The first term of Eq.~(\ref{Omeg_q}) gives  renormalized fermion 
vacuum loop correction at one-loop order ~\cite{Quiros:1999jp,Skokov:2010sf,Trvk} as : 
\begin{equation}
\Omega_{q\bar{q}}^{\rm vac} =  -\sum_{f=u,d,s} \frac{N_c}{8 \pi^2} m_f^4  \ln\left(\frac{m_f}{\text{M}}\right)
\label{Omega_reg} 
\end{equation}
The Polyakov loop potential and the temperature dependent part of the quark-antiquark contribution
to the grand potential in Eq.(\ref{eq:grandp}) vanishes at $T=0$ and $\mu=0$. 
Now the grand potential in vacuum becomes the 
renormalization scale M dependent as \cite{Trvk} :
\begin{equation}
\Omega^{\rm M} (\sigma_x,\sigma_y) =U(\sigma_x,\sigma_y)+\Omega_{q\bar{q}}^{\rm vac}
\label{Omeg_rel}
\end{equation}

The logarithmic M dependence
of the term $\Omega_{q\bar{q}}^{\rm vac}$ generates a renormalization scale M dependent part 
$\lambda_{2\text{M}}$ in the expression of the parameter $\lambda_2=\lambda_{2s}+n+\lambda_{2+}+\lambda_{\text{2M}}$. 
$\lambda_{2s}$ is the same old $\lambda_2$ parameter of the QM/PQM model in Ref.~\cite{Rischke:00,Schaefer:09,gupta}.
Here, $n=\frac{N_cg^4}{32\pi^2}$, $\lambda_{2+}=\frac{n{f_{\pi}}^2}{f_K \l f_K-f_{\pi}\r}\log\{\frac{2 f_K-f_{\pi}}{f_{\pi}}\}$
and $\lambda_{\text{2M}}= 4n\log\{ \frac{g\l 2f_K-f_{\pi}\r}{2 \text{M}}\}$.
After substituting this value of $\lambda_2$ in the expression of U($\sigma_x,\sigma_y$) 
and rearranging the terms, we find that
the logarithmic M dependence of $\lambda_2$ contained in $\lambda_{\text{2M}}$, completely cancels 
the scale dependence of all the terms in $\Omega_{q\bar{q}}^{\rm vac}$. The vacuum 
effective potential now becomes free of any renormalization scale dependence. It is expressed as 
\cite{Trvk}
\begin{eqnarray}
&& \Omega(\sigma_{x},\sigma_{y}) = \frac{m^{2}}{2}\l\sigma_{x}^{2} +
  \sigma_{y}^{2}\r -h_{x} \sigma_{x} -h_{y} \sigma_{y}
 - \frac{c}{2 \sqrt{2}} \sigma_{x}^2 \sigma_{y} \nn \\
 && + \frac{\lambda_{1}}{2} \sigma_{x}^{2} \sigma_{y}^{2}+
  \frac{\lambda_{1}}{4} \l \sigma_{x}^{4} +\sigma_{y}^{4} \r
    + \frac{ \l \lambda_{2\text{v}}+n\r}{8}\l \sigma_{x}^{4} + 2 \sigma_{y}^{4} \r \nn \\
 && -\frac{n \sigma_{x}^4}{2} \log\l \frac{\sigma_{x}}{\l2f_K-f_{\pi}\r}\r - n \sigma_{y}^4
    \log\l \frac{\sqrt 2\ \sigma_{y}}{\l2f_K-f_{\pi}\r}\r 
\label{eq:mesVop}
\end{eqnarray}
here, $\lambda_{2\text{v}}=\lambda_{2s}+\lambda_{2+}$.
The calculation of vacuum meson masses from the effective potential also shows that the scale M dependence 
completely cancels out from their expressions. The explicit derivations of scale independent meson masses 
are given in the appendix B of Ref. \cite{Trvk}. Again when $c$ becomes $c(T)$, $\Omega(\sigma_{x},\sigma_{y})$
becomes $\Omega(\sigma_{x},\sigma_{y},T)$.
			
  Now the thermodynamic grand potential in the Polyakov Quark Meson Model with 
fermion vacuum correction term (PQMVT) will be written as 
\begin{eqnarray}
\Omega_{\rm MF}(T,\mu_{f};\sigma_x,\sigma_y,\Phi,\bar\Phi) & =& {\cal
U_{\text{glue}}}(T;\Phi,\bar\Phi) + \Omega(\sigma_x,\sigma_y,T)  \nn\\
&&+\Omega_{q\bar{q}}^{\rm T}(T,\mu_{f};\sigma_x, \sigma_y, \Phi,\bar\Phi)
\label{OmegaMFPQMVT}
\end{eqnarray}
One can get the quark condensates $\sigma_x$, $\sigma_y$ and Polyakov
loop expectation values $\Phi$, $\bar\Phi$ by searching the  
minimum of the grand potential for a given value of temperature $T$ 
and chemical potential $\mu_{f}$.
\begin{equation}
\label{eq:gapeq}
  \left.\frac{ \partial \Omega_{\rm MF}}{\partial
      \sigma_x} = \frac{ \partial \Omega_{\rm MF}}{\partial \sigma_y} 
      = \frac{ \partial \Omega_{\rm MF}}{\partial \Phi}
      = \frac{\partial \Omega_{\rm MF}}{\partial \bar\Phi}
  \right|_{\sigma_x = \bar\sigma_x, \sigma_y=\bar\sigma_y, 
   \Phi=\bar\Phi} = 0\ .
\end{equation} 

%%%%%%%%%%%%%%%%%%%%%%%%%%%%%%%%%%%%%%%%%%%%%%%%%%%%%%%%%%%%%%%%%%%%%%%%%%%%%%
%%  Meson Masses and Mixing Angles
%%
%%%%%%%%%%%%%%%%%%%%%%%%%%%%%%%%%%%%%%%%%%%%%%%%%%%%%%%%%%%%%%%%%%%%%%%%%%%%%%
\section{Meson Masses and Mixing Angles}
\label{sec:mixang}

	The curvature of the grand potential in Eq.(\ref{eq:grandp}) at the 
minimum gives the finite temperature scalar and pseudo scalar meson mass matrices.
\begin{equation}
\label{eq:sdergrand}
   {m_{\alpha,ab}^{2}} \bigg|_{T}=\frac{\partial^2 \Omega_{\rm MF} (T,\mu_{f};\sigma_x, \sigma_y, \Phi,\bar\Phi)}{\partial \xi_{\alpha,a} 
 \partial \xi_{\alpha,b}} \bigg|_{min}
\end{equation} 
\begin{table*}[!hbt]
\begin{tabular}{|l|l||l|l|}
 \hline
    & Meson masses calculated from pure mesonic potential  & & Fermionic vacuum correction in meson masses \\\hline
    $(m^{\text{m}}_{a_{0}})^2$ & $m^2 +\lambda_1(x^2+y^2)+\frac{3\lambda_2}{2} x^2+
    \frac{\sqrt{2}c}{2}y $\qquad & $({\delta m^{\text{v}}_{s,11}})^{2}$ & $-\frac{N_cg^4}{64\pi^2}x^2(4+3X)$
    \\ %\hline
    $(m^{\text{m}}_{\kappa})^{2}$&$m^2+\lambda_1(x^2+y^2)+\frac{\lambda_{2}}{2}(x^2+\sqrt{2}xy+2y^2)+\frac{c}{2} x $\qquad & $({\delta m^{\text{v}}_{s,44}})^{2}$
    & $-\frac{N_cg^4}{64\pi^2}\l\frac{x+\sqrt{2}y}{x^2-2y^2}\r \l x^3X-2\sqrt{2}y^3Y\r$
    \\
    $(m^{\text{m}}_{s,00})^2$ &  $m^2+\frac{\lambda_1}{3}(7x^2+4\sqrt{2}xy+5y^2)+\lambda_2(x^2 + y^2)-\frac{\sqrt{2}c}{3} (\sqrt{2} x +y)$ &
    $({\delta m^{\text{v}}_{s,00}})^{2}$&$-\frac{N_cg^4}{96\pi^2}\l3\l x^2X+y^2Y\r+4\l x^2+y^2\r\r$ \\
    $(m^{\text{m}}_{s,88})^{2}$ & $m^2 +\frac{\lambda_1}{3}(5x^2-4\sqrt{2}xy +7y^2)+\lambda_2(\frac{x^2}{2} +2y^2)+\frac{\sqrt{2}c}{3} (\sqrt{2}x-\frac{y}{2})$ & 
    $({\delta m^{\text{v}}_{s,88}})^{2}$&  
    $-\frac{N_cg^4}{96\pi^2}\l \frac{3}{2}\l x^2X+4y^2Y\r+2\l x^2+4y^2\r\r$
    \\
   $(m^{\text{m}}_{s,08})^{2}$ &  $\frac{2\lambda_1}{3}(\sqrt{2}x^2 -xy -\sqrt{2}y^2) +\sqrt{2}\lambda_2(\frac{x^2}{2}-y^2) +\frac{c}{3\sqrt{2}}(x- \sqrt{2}y)$ &
  $({\delta m^{\text{v}}_{s,08}})^{2}$&$-\frac{N_cg^4}{8\sqrt{2}\pi^2}\l\frac{1}{4}\l x^2X-2y^2Y\r+\frac{1}{3}\l x^2-2y^2\r\r$   
   \\
$ (m^{\text{m}}_{\pi})^{2}$ & $m^2 + \lambda_1 (x^2 + y^2) +\frac{\lambda_2}{2} x^2 -\frac{\sqrt{2} c}{2} y$& $({\delta m^{\text{v}}_{p,11}})^{2}$  
&$-\frac{N_cg^4}{64\pi^2}x^2X$  \\ %\hline
$(m^{\text{m}}_{K})^{2}$ &$m^2 + \lambda_1 (x^2 + y^2) +\frac{\lambda_2}{2} (x^2 - \sqrt{2} x y +2 y^2) - \frac{c}{2} x$&$({\delta m^{\text{v}}_{p,44}})^{2}$
&$-\frac{N_cg^4}{64\pi^2}\l\frac{x-\sqrt{2}y}{x^2-2y^2}\r \l x^3X+2\sqrt{2}y^3Y\r$  \\ 
$(m^{\text{m}}_{p,00})^{2}$ & $m^2 + \lambda_1(x^2 +y^2) + \frac{\lambda_2}{3}(x^2 +y^2) + \frac{c}{3} (2x + \sqrt{2} y)$&$({\delta m^{\text{v}}_{p,00}})^{2}$
&$-\frac{N_cg^4}{96\pi^2}\l x^2X+y^2Y\r$ \\
$(m^{\text{m}}_{p,88})^{2}$ & $m^2 +\lambda_1(x^2 +y^2) +\frac{\lambda_2}{6}(x^2 +4y^2)-\frac{c}{6}(4x -\sqrt{2}y)$& $({\delta m^{\text{v}}_{p,88}})^{2}$
&$-\frac{N_cg^4}{192\pi^2}\l x^2X+4y^2Y\r$ \\
$(m^{\text{m}}_{p,08})^{2}$ & $\frac{\sqrt{2}\lambda_2}{6}(x^2-2y^2)-\frac{c}{6}(\sqrt{2}x -2y)$& $({\delta m^{\text{v}}_{p,08}})^{2}$
&$-\frac{N_cg^4}{96\sqrt{2}\pi^2}\l x^2X-2y^2Y\r$ \\ 
\hline
\end{tabular}
\caption{Expressions for $({m^{\text{m}}_{\alpha,ab}})^{2}$ in the left half of the table are the vacuum
meson masses calculated from the second derivatives of the pure mesonic potential U($\sigma_x,\sigma_y$).
Evaluated expressions of mass modifications $({\delta m^{\text{v}}_{\alpha,ab}})^{2}$ due to the 
fermionic vacuum correction are given in the right half. Symbols used in the expressions are defined as $x=\sigma_x$, 
$y=\sigma_y$, $X=\l1+4\log\l \frac{g\sigma_x}{2\text{M}}\r\r$ and 
$Y=\l1+4\log\l \frac{g\sigma_y}{\sqrt{2}\text{M}}\r\r$.Here, when $\alpha$ =s, the (11)
element gives squared mass of the scalar $a_{0}$ meson which is degenerate with the (22) and (33) elements.
Similarly, the (44) element which is degenerate with (55), (66) and (77) elements, gives 
the squared $\kappa$ meson mass. The squared  $\sigma$ and $f_{0}$ meson masses are obtained by diagonalizing the scalar mass matrix in
the (00)-(88) sector and we get a scalar mixing angle $\theta_{S}$. We have a completely analogous situation for the pseudo scalar sector
($\alpha$ = p)  with the following identification; the (11) element gives the squared pion mass and the squared kaon mass
is given by the (44) element.}
\label{tab*:MesonVM}
\end{table*}
The subscript $\alpha=$ s, p ; s stands for scalar and p stands 
for pseudo scalar mesons and $a, b=0 \cdots 8$. 
\begin{equation}
\label{eq:MasFT}
{m_{\alpha,ab}^{2}} \bigg|_{T} =({m^{\text{m}}_{\alpha,ab}})^{2}+({\delta m^{\text{v}}_{\alpha,ab}})^{2}
+ ({\delta m^{\text{T}}_{\alpha,ab}})^{2}
\end{equation}
Where  $({m^{\text{m}}_{\alpha,ab}})^{2}$ is the tree level mass matrix,  $({\delta m^{\text{v}}_{\alpha,ab}})^{2} $ and 
$({\delta m^{\text{T}}_{\alpha,ab}})^{2}$ are the contributions from fermionic vacuum and thermal fluctuations, respectively.

The mass modification in vacuum ($T=0$, $\mu=0$), due to the fermionic vacuum correction will be given by
\begin{eqnarray}
&&({\delta m^{\text{v}}_{\alpha,ab}})^{2}=\frac{\partial^2 \Omega_{q\bar{q}}^{\rm vac}}
{\partial \xi_{\alpha,a}\partial \xi_{\alpha,b}} \bigg|_{min} \nn \\
&&=-\frac{N_c}{8\pi^2}\sum_f\Bigl[\l
2\log\l\frac{m_f}{\text{M}}\r+\frac{3}{2} \r\l\frac{\partial m_f^2}{\partial
 \xi_{\alpha,a}}\r\l\frac{\partial m_f^2}{\partial \xi_{\alpha,b}}\r \nn \\
&&+\l\frac{m_f^2}{2}+2m_f^2\log\l\frac{m_f}{\text{M}} \r \r \frac{\partial^2
  m_f^2}{\partial \xi_{\alpha,a}\partial \xi_{\alpha,b}}\Bigr]
\label{eq:dFVac}
\end{eqnarray}

Here  $m^{2}_{f,a} \equiv \partial m^{2}_{f}/ \partial \xi_{\alpha,a}$ denotes
the first partial derivative and 
$m^{2}_{f,ab} \equiv \partial m^{2}_{f,a}/ \partial \xi_{\alpha,b}$ signifies
the second partial derivative of the squared quark mass with respect to the meson 
fields $\xi_{\alpha,b}$. These derivatives are evaluated in the non strange-strange
basis  and given in the Table III of Ref.\cite{Schaefer:09}. 
The mass modifications given in Eq.(\ref{eq:dFVac}) have been evaluated \cite{Trvk, Sandeep}
and different expressions of $({\delta m^{\text{v}}_{\alpha,ab}})^{2}$ are collected in the Table \ref{tab*:MesonVM}
for all the mesons of scalar and pseudo-scalar nonet. The vacuum mass expressions $({m^{\text{m}}_{\alpha,ab}})^{2}$
for these mesons as originally evaluated from the second derivative of pure mesonic potential in 
Ref.~\cite{Rischke:00,Schaefer:09}, are also given in this Table. Further the diagonalization of the (00)-(88) sector 
of the pseudo scalar mass matrix, gives 
the squared masses of $\eta$ and  $\eta'$ and analogously, we get a pseudo-scalar mixing angle $\theta_{P}$. 
The mass expressions $({m^{\text{m}}_{\alpha,ab}})^{2}$ have a renormalization scale M dependence in the 
PQMVT model due to the parameter $\lambda_{2}$. This dependence gets completely canceled by the 
already existing scale M dependence in the mass modifications $({\delta m^{\text{v}}_{\alpha,ab}})^{2} $ and
the final expressions of vacuum meson masses $m_{\alpha,ab}^{2}$ which is equal to 
$({m^{\text{m}}_{\alpha,ab}})^{2}+({\delta m^{\text{v}}_{\alpha,ab}})^{2}$ ,
are free of any renormalization scale 
dependence as shown explicitly in the appendix B of Ref.\cite{Trvk} .

The correction to the meson mass matrix due to the fermionic thermal fluctuations in the 
presence of Polyakov loop as originally evaluated in Ref.\cite{gupta}
in the PQM model is written as:
\begin{eqnarray}
\label{eq:ftmass}
&&({\delta m^{\text{T}}_{\alpha,ab}})^{2} \Big|_{PQM} = \frac{\partial^2\Omega_{q\bar{q}}^{\rm T} (T,\mu_{f}, \sigma_x, \sigma_y, \Phi, \bar\Phi)}
 {\partial \xi_{\alpha,a} \partial \xi_{\alpha,b}} \Big|_{min} \nn \\
&&= 3 \sum_{f=x,y} \int \frac{d^3 p}{(2\pi)^3} \frac{1}{E_f}
 \biggl[ (A^{+}_{f} + A^{-}_{f} ) \biggl( m^{2}_{f,a b} - \frac{m^{2}_{f,a}
 m^{2}_{f, b}}{2 E_{f}^{2}} \biggl)  \nn \\
      && + (B_{f}^{+} + B_{f}^{-}) \biggl(  \frac{m^{2}_{f,a}  m^{2}_{f, b}}{2 E_{f} T}
\biggl) \biggl]  
\end{eqnarray}
The notations $A_{f}^{\pm}$ and $B_{f}^{\pm}$ have the following 
definitions
\begin{equation}
A_{f}^{+} =  \frac{\Phi e^{-E_{f}^{+}/T} + 2 \bar\Phi e^{-2E_{f}^{+}/T} + e^{-3E_{f}^{+}/T}}{g_{f}^{+}}
\end{equation}
\begin{equation}
A_{f}^{-} =  \frac{\bar\Phi e^{-E_{f}^{-}/T} + 2 \Phi e^{-2E_{f}^{-}/T} + e^{-3E_{f}^{-}/T}}{g_{f}^{-}}
\end{equation}

and $B_{f}^{\pm}={3(A_{f}^{\pm}})^2 - C_{f}^{\pm}$, where we 
again define 

\begin{equation}
C_{f}^{+} = \frac{\Phi e^{-E_{f}^{+}/T} + 4 \bar\Phi e^{-2E_{f}^{+}/T} +3 e^{-3E_{f}^{+}/T}}{g_{f}^{+}} 
\end{equation}
\begin{equation}
C_{f}^{-} = \frac{\bar\Phi e^{-E_{f}^{-}/T} + 4 \Phi e^{-2E_{f}^{-}/T} +3 e^{-3E_{f}^{-}/T}}{g_{f}^{-}} 
\end{equation}
In the present work we are 
considering symmetric quark matter and net baryon number to be zero. 
  
The diagonalization of (0,8) component 
of mass matrix gives the masses of $\sigma$ and $f_0$ mesons in scalar 
sector and the masses of $\eta'$ and $\eta$ in pseudo scalar sector. 
The scalar mixing angle $\theta_s$ and pseudo scalar mixing angle 
$\theta_p$ are given by
\begin{equation} 
\tan{2\theta_{\alpha}} = \biggl( \frac{2 m^2_{\alpha,08}}
{m^2_{\alpha,00}-m^2_{\alpha,88}} \biggr)
\end{equation}

The appendix C of Ref.\cite {Schaefer:09} contains all transformation details of the mixing for the (0,8) basis
that generates the physical basis of the scalar ($\sigma$,$f_0$) and pseudo-scalar ($\eta'$, $\eta$) mesons. This appendix
also explains the ideal mixing, and gives the details of formulae by which the physical mesons transform into the
pure strange or non-strange quark system of mesons.

\section{Model Parameter Fixing}
\label{sec:parameter}
The six unknown parameters $\text{m}^2$, $\lambda_1,\lambda_2$, $h_x,h_y$ and $c$ in the mesonic 
potential U($\sigma_x,\sigma_y$), are fixed in the vacuum by six experimentally known quantities. The 
known values of $m_\pi $, $m_K $, the average squared mass of $\eta$ and $\eta^{\prime}$ mesons,
$m_{\eta}^{2}+m_{\eta^{\prime}}^{2} $, and the pion and kaon decay constants $f_{\pi}$ and $f_{K}$ and finally
the $m_{\sigma} $ are used to obtain the value six parameters in vacuum \cite{Rischke:00, Schaefer:09}. 
\begin{table*}[!hbt]
  % \caption{The squared masses of scalar and pseudoscalar mesons appear 
  % in nonstrange strange basis. In the following table x denotes $\sigma_x$ 
  % and y denotes $\sigma_y$. The  masses of nonstrange $\sigma_{NS}$, 
  % strange $\sigma_S$, nonstrange $\eta_{NS}$ and strange $\eta_S$  mesons 
  % are given in the last two rows.}
  \begin{tabular}{|l|l||l|l|} 
    \hline
    & Scalar Meson Masses & & Pseudo scalar Meson Masses \\\hline
    $m^{2}_{\sigma}$ & $m^2_{s,{00}}\cos^{2}\theta_{s}+m^2_{s,88}\sin^{2}\theta_{s}+2m^{2}_{s,08}\sin\theta_{s}\cos\theta_{s}$
    & $m^{2}_{\eta'}$ & $m^2_{p,{00}}\cos^{2}\theta_{p}+m^2_{p,88}\sin^{2}\theta_{p}+2m^{2}_{p,08}\sin\theta_{p}\cos\theta_{p}$ \\
    $m^{2}_{f_0}$ & $m^2_{s,00}\sin^{2}\theta_{s}+m^{2}_{s,88}\cos^{2}\theta_{s}-2m^2_{s,08}\sin\theta_{s}\cos\theta_{s}$\qquad
    & $m^{2}_{\eta}$ & $m^2_{p,00}\sin^{2}\theta_{p}+m^{2}_{p,88}\cos^{2}\theta_{p}-2m^2_{p,08}\sin\theta_{p}\cos\theta_{p}$\\
    $m^{2}_{\sigma_{NS}}$ & $\frac{1}{3}(2m^{2}_{s,00}+m^{2}_{s,88}+2\sqrt{2}m^{2}_{s,08})$ &
    $m^{2}_{\eta_{NS}}$ & $\frac{1}{3}(2m^{2}_{p,00}+m^{2}_{p,88}+2\sqrt{2}m^{2}_{p,08})$\\
    $m^{2}_{\sigma_{S}}$ & $\frac{1}{3}(m^{2}_{s,00}+2m^{2}_{s,88}-2\sqrt{2}m^{2}_{s,08})$ &
    $m^{2}_{\eta_{S}}$ & $\frac{1}{3}(m^{2}_{p,00}+2m^{2}_{p,88}-2\sqrt{2}m^{2}_{p,08})$\\
    \hline
  \end{tabular}
 \caption{The squared masses of scalar and pseudo scalar mesons which are obtained after the diagonalization of the 00-88 sector of
  mass matrix. The meson masses in the non strange $\sigma_{NS}(\eta_{NS})$ and strange $\sigma_S(\eta_S)$ basis are given in the
  last two rows.}
  \label{tab:mass}
\end{table*}

\begin{table*}[!ht]
%\caption{parameters for $m_{\sigma} = 600$ MeV with and without $U_A(1)$ 
%axial anomaly term.}
 \begin{tabular}{|l|l|l|l|l|l|l|l|l|l|}
\hline
$m_\sigma [MeV]$ &$T_{glue}^{c} [MeV]$ & $c [MeV]$  & $m^2 [MeV^2]$ & $\lambda_1$ & $\lambda_{2s}$ & $h_x [MeV^3]$& $h_y [MeV^3]$ & $T_1[MeV]$ & $b_1[MeV]$ \\ 
\hline
$400$ &$240$ & $4807.84$ & $(282.67)^2$  &  $-8.18$   & $46.48$ & $(120.73)^3$ &$(336.41)^3$  & $108$ & $90$\\ 
$500$ &$240$ & $4807.84$ & $(171.73)^2$ &$-5.29$ & $46.48$ & $(120.73)^3$ &$(336.41)^3$   & $105$ & $78$\\ 
$600$ & $240$ & $4807.84$ &$-(183.73)^2$ & $ -1.689$   & $46.48$ & $(120.73)^3$ & $(336.41)^3$  & $105$ & $62$ \\ \hline 

 \end{tabular}
\caption{Parameters for different values of sigma meson mass $m_{\sigma}$. We have chosen a combination of sigma meson mass and 
Polyakov loop potential parameter ($m_{\sigma},T_{glue}^{c}$) and found for this combination, the values of the parameters 
$T_1$ and $b_1$ which give the temperature dependence of KMT coupling $c(T)$ by obtaining the most suitable coincidence of the 
temperature variations of the subtracted chiral condensate $\Delta_{l,s} (T) $ calculated in our PQMVT-II model with the 
Wuppertal group lattice QCD data for the same.}
\label{tab:pmtr}
\end{table*}

In the PQMVT model calculations, the vacuum mass expressions that determine $\lambda_2$ and c are
\begin{widetext}

$m^{2}_{\pi}=({m^{\text{m}}_{\pi}})^{2}+({\delta m^{\text{v}}_{p,11}})^{2}$, 
$m^{2}_K=({m^{\text{m}}_K})^{2}+({\delta m^{\text{v}}_{p,44}})^{2}$ and $m^{2}_{\eta}+m^{2}_{\eta'}=m^{2}_{p,00}+m^{2}_{p,88}$ where
$m^{2}_{p,00}=({m^{\text{m}}_{p,00}})^{2}+({\delta m^{\text{v}}_{p,00}})^{2}$ and 
$m^{2}_{p,88}=({m^{\text{m}}_{p,88}})^{2}+({\delta m^{\text{v}}_{p,88}})^{2}$. We can write $m^{2}_{\eta}+m^{2}_{\eta'}=
({m^{\text{m}}_{\eta}})^{2}+({m^{\text{m}}_{\eta'}})^{2}+({\delta m^{\text{v}}_{p,00}})^{2}+({\delta m^{\text{v}}_{p,88}})^{2}$ 
where $({m^{\text{m}}_{\eta}})^{2}+({m^{\text{m}}_{\eta'}})^{2}=
({m^{\text{m}}_{p,00}})^{2}+({m^{\text{m}}_{p,88}})^{2}$. Using mass modification  expressions
$({\delta m^{\text{v}}_{\alpha,ab}})^{2}$  given in the Table \ref{tab*:MesonVM}, we write

\begin{eqnarray}
&&({m^{\text{m}}_K})^{2}=m^{2}_K+\frac{N_cg^4}{64\pi^2}\l\frac{x-\sqrt{2}y}{x^2-2y^2}\r 
\l x^3X+2\sqrt{2}y^3Y\r\ ; \ ({m^{\text{m}}_{\pi}})^{2}=m^{2}_{\pi}+
\frac{N_cg^4}{64\pi^2}x^2X\   \nn\\ 
&& \text{and} \ \ \ \l({m^{\text{m}}_{\eta}})^{2}+({m^{\text{m}}_{\eta'}})^{2}\r=\l m^{2}_{\eta}+m^{2}_{\eta'}\r+
\frac{N_cg^4}{192\pi^2}\l3x^2X+6y^2Y\r\
\label{PIKETA}
\end{eqnarray}
The $f_{\pi}$ and $f_K$ give vacuum condensates according to the partially conserved axial vector current (PCAC) relation. The
x =$\sigma_x=f_{\pi}$ and y =$\sigma_y=\l \frac{2f_K-f_{\pi}}{\sqrt{2}}\r$ at $T=0$. The parameters $\lambda_2$ and c  
in vacuum are obtained as:
\begin{eqnarray}
\lambda_2&=&\frac{3\l2f_K-f_{\pi}\r({m^{\text{m}}_K})^{2}-\l2f_K+f_{\pi}\r({m^{\text{m}}_{\pi}})^{2}
-2\l({m^{\text{m}}_{\eta}})^{2}+({m^{\text{m}}_{\eta'}})^{2}\r 
\l f_K-f_{\pi}\r}{\l3f^2_{\pi}+8f_K\l f_K-f_{\pi}\r\r\l f_K-f_{\pi}\r}\ 
\label{lam2c} 
\end{eqnarray}
\begin{equation}
c=\frac{({m^{\text{m}}_K})^{2}-({m^{\text{m}}_{\pi}})^{2}}{f_K-f_{\pi}}-\lambda_2\l2f_K-f_{\pi}\r
\label{eqc}
\end{equation}
When expressions of $({m^{\text{m}}_{\pi}})^2$, $({m^{\text{m}}_K})^2$ and $\l ({m^{\text{m}}_{\eta}})^2+
({m^{\text{m}}_{\eta'}})^2\r$ from Eq.(\ref{PIKETA}) are substituted in Eq.(\ref{lam2c}) and Eq.(\ref{eqc}) 
and the vacuum value of the condensates are used, the final rearrangement of terms yields:
\begin{eqnarray}
%\eqalign
\label{lam2VT}
\lambda_2&=&\lambda_{2s}+n+\lambda_{2+}+\lambda_{\text{2M}}\ \text{where} \ \lambda_{2s}=\frac{3\l2f_K-f_{\pi}\r m^{2}_K-\l2f_K+f_{\pi}\r m^{2}_{\pi}-2\l m^{2}_{\eta}+m^{2}_{\eta'}\r \l f_K-f_{\pi}\r}{\l3f^2_{\pi}+8f_K\l f_K-f_{\pi}\r\r\l f_K-f_{\pi}\r}\ ,\nn\\
n&=&\frac{N_cg^4}{32\pi^2}, 
\ \lambda_{2+}=\frac{n{f_{\pi}}^2}{f_K \l f_K-f_{\pi}\r}\log\l 
\frac{2\ f_K-f_{\pi}}{f_{\pi}}\r
\text{and scale dependent part} \ \lambda_{\text{2M}}= 4n\log\l \frac{g\l 2f_K-f_{\pi}\r}{2 \text{M}}\r
\end{eqnarray}
\begin{equation}
\label{cfinal}
c=\frac{{m_K}^{2}-{m_{\pi}}^{2}}{f_K-f_{\pi}}-\lambda_{2s}\l2f_K-f_{\pi}\r
\end{equation}
\end{widetext}
\sloppy
We note that the $\lambda_{2s}$ is equal to the earlier
$\lambda_2$ parameter determined in the QM/PQM model calculations in 
Ref.~\cite{Rischke:00,Schaefer:09,gupta}. We have used the numerical values $f_{\pi}=92.4$ \ MeV, $f_{K}=113.0$ \ MeV, 
$m_{\pi}=138.0$ \ MeV, $m_{K}=496.0$ \ MeV,$m_{\eta}=539.0$ \ MeV and $m_{\eta^{\prime}}=963.0$ \ MeV .
In the present calculation, the proper renormalization 
of fermionic vacuum, leads to the augmentation of $\lambda_{2s}$ by the
addition of a term (n+$\lambda_{2+}$) and further, we get a renormalization scale M dependent 
contribution $\lambda_{\text{2M}}$ in the expression of the $\lambda_2$ in Eq.(\ref{lam2VT}).

We get the complete cancellation of M dependence in the evaluation of $c$ also and finally 
its value turns out to be the same as in the QM model. 
The scale M independent expression of $m^2_{\pi}$ obtained in the appendix B of Ref.\cite{Trvk} can be used with 
x=$f_{\pi}$ and y=$(\frac{2f_K-f_{\pi}}{\sqrt{2}})$, to express $m^2$ in terms of $\lambda_1$.
\begin{eqnarray}
m^2&=&m^{2}_{\pi}-\lambda_1\{f_{\pi}^2+\frac{(2f_K-f_{\pi})^2}{2}\}-\frac{f_{\pi}^2}{2} \ \biggl[\ \lambda_{2\text{v}}-\ 4 \ n\nn\\
&& \log \ \{\ \frac{f_{\pi}}{\l2f_K-f_{\pi}\r} \}\biggl] +\frac{c}{2} \ (\ 2f_K-f_{\pi})
\end{eqnarray}
When we use the formula of $m^2_{\sigma}$ in the Table \ref{tab:mass} of the appendix B of Ref\cite{Trvk} 
(with the vacuum values of the masses $m^2_{s,00}$, $m^2_{s,88}$, $m^2_{s,08}$ and the mixing angle $\theta_S$) and 
substitute the above expression of $m^2$ in it, we will get the numerical value of $\lambda_1$ for different values of
$m_\sigma$ when we put $m_\pi$=138 MeV . The explicit symmetry breaking parameters $h_x$ 
and $h_y $ in the nonstrange-strange basis are related to the pion and kaon masses by the ward 
identities \cite{Rischke:00, Schaefer:09}.
\begin{eqnarray}
h_{x}=f_{\pi} m_{\pi}^2 \ \ \ \  , h_{y}=\sqrt{2} f_{K} m_{K}^2-\frac{f_{\pi} m_{\pi}^2}{\sqrt{2}}
\end{eqnarray}
The last parameter , the Yukawa coupling g, is fixed from the nonstrange constituent quark mass 
\begin{equation}
g=\frac{2m_{q}}{\sigma_{x}}
\end{equation}
The  Yukawa coupling value $g= 6.5$ has been fixed from the non strange 
constituent quark mass $m_q = 300$ MeV in vacuum ($T=0, \mu=0)$. This predicts the 
vacuum strange quark mass $m_s \simeq 433$ MeV.
We have chosen a combination of sigma meson mass and 
Polyakov loop potential parameter ($m_{\sigma},T_{glue}^{c}$) and found for this combination, the values of the parameters 
$T_1$ and $b_1$ which give the temperature dependence of KMT coupling $c(T)$ by obtaining the most suitable coincidence of the 
temperature variations of the subtracted chiral condensate $\Delta_{l,s} (T) $ calculated in our PQMVT-II model with the 
Wuppertal group lattice QCD data for the same. Fixing $T_{glue}^{c}$ at 240 MeV, we obtain a good coincidence of 
$\Delta_{l,s} (T) $ temperature  variation with the LQCD data when temperature dependence of $c(T)$ in PQMVT-II model becomes 
steep and steeper with small and smaller values of the parameters $T_1$ and $b_1$ for large and larger values of the sigma meson 
mass $m_{\sigma}$. 
 
%%%%%%%%%%%%%%%%%%%%%%%%%%%%%%%%%%%%%%%%%%%%%%%%%%%%%%%%%%%%%%%%%%%%%%%%%%%%%%
%%%%%%%%%%%%%%%%%%%%%%%%%%%%%%%%%%%%%%%%%%%%%%%%%%%%%%%%%\langle%%%%%%%%%%%%%%%%%%%%%
\section{Temperature Dependence of c(T) and Chiral  Restoration}
\label{sec:eploop}

We are exploring, how the temperature dependence of $ U_{A}(1) $ breaking coupling strength
c(T) in the $2+1$ flavor Polyakov quark meson model with the fermionic vacuum 
correction term (PQMVT), influences the chiral symmetry restoration and the axial 
$ U_{A}(1) $ symmetry restoration. For comparing the results in scenario one,  when $ U_{A}(1) $ 
breaking coupling strength c is constant, we call the PQMVT model as PQMVT-I while in scenario 
two, when $ U_{A}(1) $ breaking coupling strength c(T) is temperature dependent, we call PQMVT model 
as PQMVT-II. The temperature variations of nonstrange condensate, strange condensate and Polyakov loop expectation value 
are quite sensitive to the sigma meson mass in both the model scenarios. Table \ref{tab:pmtr}, shows 
values of all the parameters where $\lambda_1$ , $m^2$ change significantly depending on the $m_\sigma$
value. We have fitted the Wuppertal group LQCD data for the subtracted chiral order parameter \cite{LQCDWB2} with the corresponding
$\Delta_{ls}$ temperature variation in the PQMVT-II model for fixed value of $T_{glue}^{c}$= 240 MeV and three different
values of $m_\sigma$. When $m_\sigma$=400 MeV, the parameter set $T_1$=108 MeV and $b_1$=90 MeV for the temperature 
dependence of $c(T)$ in the PQMVT-II model calculation, gives a good coincidence of $\Delta_{ls}$ temperature variation
with the LQCD data. For $m_\sigma$=500 MeV, we obtain a good fit to the LQCD data when the temperature dependence of 
$c(T)$ becomes steeper with smaller $T_1$=105 MeV and $b_1$=78 MeV. For larger  $m_\sigma$ = 600 MeV, in order to find a  
good fit of the $\Delta_{ls}$ temperature variation with the LQCD data,
the temperature dependence of $c(T)$ becomes steepmost with still smaller
$T_1$=105 MeV and $b_1$=62 MeV in the PQMVT-II model.

%%%%%%%%%%%%%%%%%%%%%%%%%%%%%%%%%%%%%%%%%%%%%%%%%%%%%%%%%%%%%%%%%%%%%%%%%%%%%%
\subsection{Effect of c(T) on Condensates }
%%%%%%%%%%%%%%%%%%%%%%%%%%%%%%%%%%%%%%%%%%%%%%%%%%%%%%%%%%%%%%%%%%%%%%%%%%%%%%
% Table for Critical Temperatures
%%%%%%%%%%%%%%%%%%%%%%%%%%%%%%%%%%%%%%%%%%%%%%%%%%%%%%%%%%%%%%%%%%%%%%%%%%%%%%
%%%%%%%%%%%%%%%%%%%%%%%%%%%%%%%%%%%%%%%%%%%%%%%%%%%%%%%%%%%%%%%%%%%%%%%%%%%%%%
%% Figure 1(a) and 1(b) in the 
%presence of axial $U_A(1)$ breaking term,in the absence of the $U_A(1)$ 
%axial breaking term,
%%%%%%%%%%%%%%%%%%%%%%%%%%%%%%%%%%%%%%%%%%%%%%%%%%%%%%%%%%%%%%%%%%%%%%%%%%%%%
\begin{table*}[!htb]
\begin{tabular}{|c|c|c|c|c|c|c|}
\hline
& PQMVT-I & PQMVT-II &PQMVT-I &PQMVT-II &PQMVT-I & PQMVT-II  \\
& $m_\sigma$=400 MeV & $m_\sigma$=400 MeV & $m_\sigma$=500 MeV & $m_\sigma$=500 MeV & $m_\sigma$=600 MeV & $m_\sigma$=600 MeV\\
& $T_{glue}^{c}$=240 MeV & $T_{glue}^{c}$=240 MeV & $T_{glue}^{c}$=240 MeV & $T_{glue}^{c}$=240 MeV & $T_{glue}^{c}$=240 MeV & $T_{glue}^{c}$=240 MeV\\ 
\hline
$T_{c}^{\chi}$(MeV)& $175.6$ & $154.9$ & $187.9$ & $158.3$ & $201.3$ & $159.0$ \\
$T_{s}^{\chi}$(MeV)&$261.7\pm2$&$265.2\pm2$&$267.9\pm3.4$ &$268.9\pm3.3$ &$274.7\pm3.5$ &$275.7\pm3.5$ \\
$T_{c}^{\Phi}$(MeV)& $155.3$ & $145.8$ & $154.3$ & $146.1$ & $155.3$ & $146.1$ \\
\hline
\end{tabular}
\caption{The table of characteristic temperature (pseudo critical temperature) 
for the chiral transition in the non-strange sector $T_{c}^{\chi}$, 
strange sector $T_{s}^{\chi}$ and the confinement-deconfinement transition $T_{c}^{\Phi}$,
in the PQMVT-I and PQMVT-II models. $\pm$ gives the temperature range near  $T_{s}^{\chi}$  over which the
rather flat and broad second peak of the strange condensate derivative $\frac{\partial\sigma_y}{\partial T}$, 
shows a distinct change of about 0.1 percent of the numerical value of the second peak height. }
\label{tab:ctemp}
\end{table*}
   On solving the gap equations Eq.(\ref{eq:gapeq})  at zero 
chemical potential, we get the temperature dependence of the Polyakov loop expectation value
$\Phi$, non strange and strange condensates . The inflection point of these order 
parameters respectively give the characteristic temperature 
(pseudo-critical temperature) for the confinement - deconfinement 
transition $T_{c}^{\Phi}$, the chiral transition in the non-strange 
$T_{c}^{\chi}$ and strange sector $T_{s}^{\chi}$. Table \ref{tab:ctemp} shows
the various pseudo-critical temperatures in two model scenarios for different values 
of $m_{\sigma}$.
We can not have direct comparison of non-strange condensate $<\sigma_{x}>$ and strange condensate
$<\sigma_{y}>$ with the lattice data . In the limit of zero light quark mass
(chiral limit), QCD has a chiral symmetry and at finite temperature, a true phase transition occurs. To eliminate quadratic divergences
in the linear quark mass dependent correction to the chiral condensate, one calculates a suitable combination of 
light and strange quark condensates at finite temperature. This quantity is further normalized by the corresponding combination of condensates
calculated at $T=0$ MeV \cite{cheng2008}. Subtracted chiral   
condensate $ \Delta_{l,s}(T) $ is such an observable which works as an order parameter
for the chiral symmetry breaking.
 It is written as  
\begin{equation}
\label{eq:Delta}
\Delta_{l,s}(T)=\frac{<\sigma_{x}>(T)-(\frac{h_{x}}{h_{y}})<\sigma_{y}>(T)}
{<\sigma_{x}>(0)-(\frac{h_{x}}{h_{y}})<\sigma_{y}>(0)}
\end{equation}
%%%%%%%%%%%%%%%%%%%%%%%%%%%%%%%%%%%%%%%%%%%%%%
\begin{figure*}[!htbp]
\subfigure[Scaled non-strange , strange condensate and $\Phi$ variation with temperature]{
\label{fig:mini:fig1:a} %% label for first subfigure
\begin{minipage}[b]{0.43\linewidth}
\centering \includegraphics[width=3.15in]{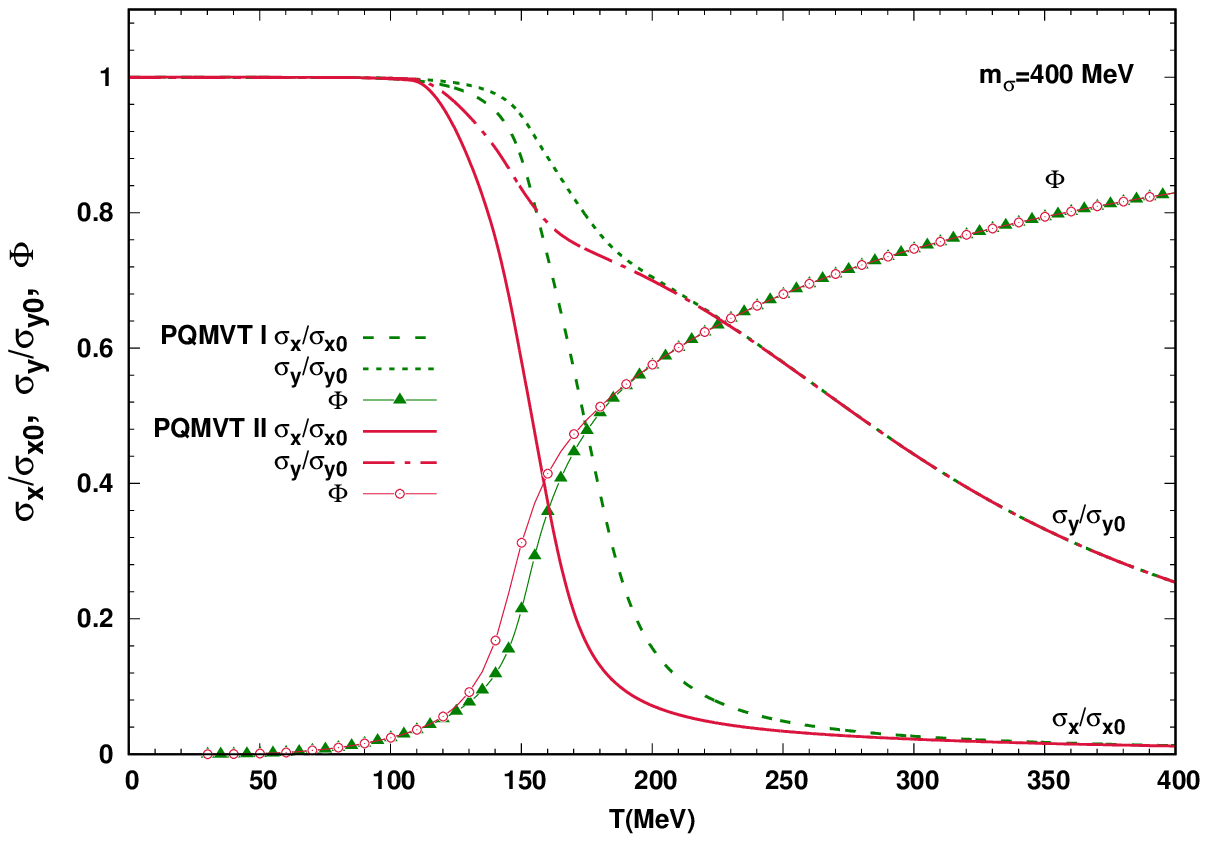}
\end{minipage}}
\hspace{.25in}
\subfigure[Subtracted chiral condensate $\Delta_{l,s}$ variation with temperature]{
\label{fig:mini:fig1:b} %% label for second subfigure
\begin{minipage}[b]{0.43\linewidth}
\centering \includegraphics[width=3.15in]{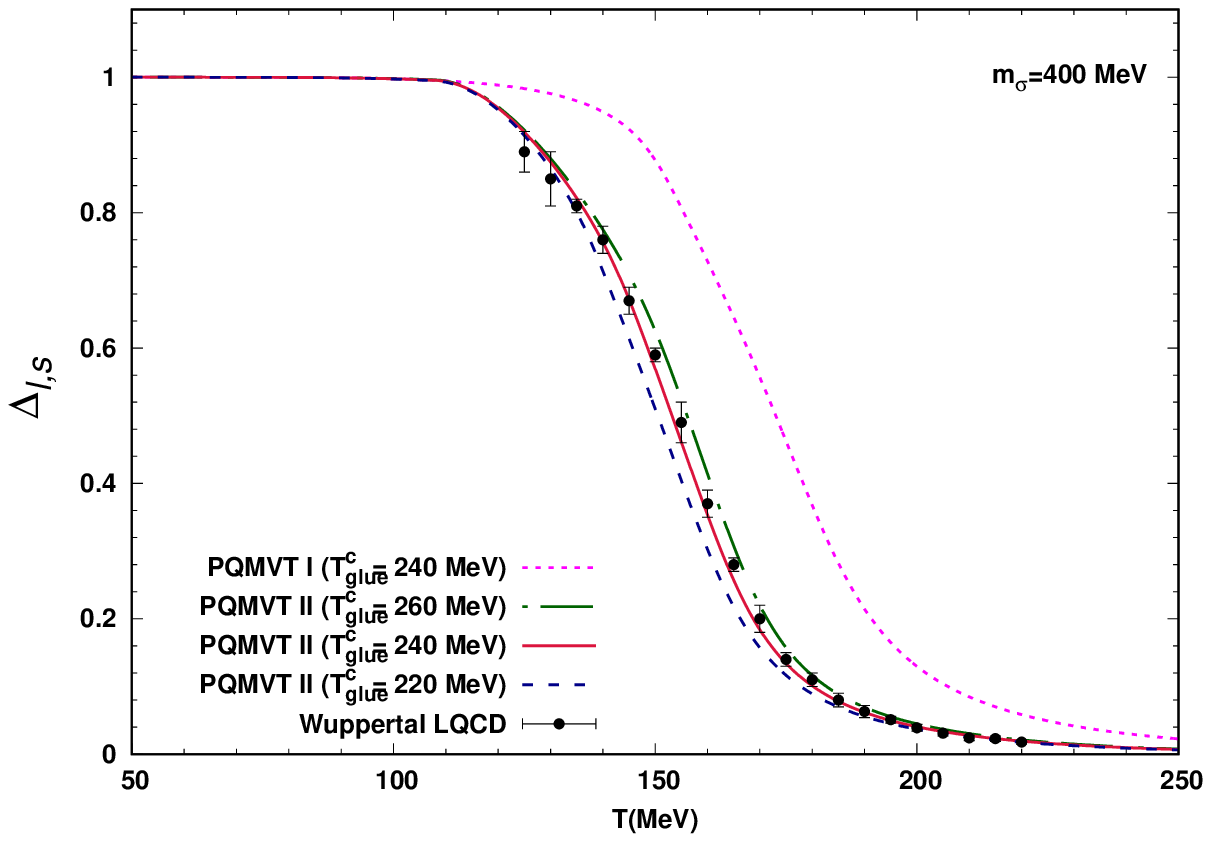}
\end{minipage}}
\caption{The Fig.\ref{fig:mini:fig1:a}, \ shows the  temperature 
variations  \ of \ the scaled non strange condensate ($\frac{\sigma_{x}}{\sigma_{x0}}$) , strange condensate 
($\frac{\sigma_{y}}{\sigma_{y0}}$ ) and the polyakov loop field $\Phi$ 
at zero chemical potential \ ($\mu = 0$), sigma meson mass $m_\sigma=400$ \ MeV and $T_{glue}^{c}=240$ MeV.
The dash line, the dotted line and the 
solid line with filled triangles in green color, represent the respective ( $\frac{\sigma_{x}}{\sigma_{x0}}$ ),
($\frac{\sigma_{y}}{\sigma_{y0}}$) and $\Phi $ temperature variations for the constant c  PQMVT-I model, while the solid line,
the dash dotted line and the solid line with filled circles in deep red color, represent the respective ( $\frac{\sigma_{x}}{\sigma_{x0}}$ ),
($\frac{\sigma_{y}}{\sigma_{y0}}$) and $\Phi $ temperature variations for the temperature dependent c(T) PQMVT-II model.
In Fig.\ref{fig:mini:fig1:b}, the dotted line in magenta represents temperature variations of $\Delta_{ls}$  for the PQMVT-I model
when $T_{glue}^{c}=240$ MeV and $m_\sigma=400$ \ MeV. The deep red solid line obtained with $T_1=108$ MeV and $b_1=90$ MeV
for the temperature dependence of c(T) in the PQMVT-II model, shows the $\Delta_{ls}$  temperature 
variations when we choose $m_\sigma=400$ \ MeV and  $T_{glue}^{c}=240$ MeV, it gives most suitable fit to
the solid black dots (except for the two upper  data points at T=125 MeV and 130 MeV) located in the middle of the error bars  
representing Wuppertal group lattice QCD data. The dash 
dotted line in green obtained with $T_1=108$ MeV and $b_1=90$ MeV, shows the other $\Delta_{ls}$  temperature 
variations when we choose $T_{glue}^{c}=260$ MeV for $m_\sigma=400$ \ MeV, it almost touches from above, the upper
error bars of the lattice QCD data points (except for the two upper data points at T=125 MeV and 130 MeV 
and lower data points after 180 MeV).
The dash line in deep blue obtained with $T_1=108$ MeV and $b_1=90$ MeV, shows an other $\Delta_{ls}$  temperature 
variations when we choose $T_{glue}^{c}=220$ MeV for $m_\sigma=400$ \ MeV, it almost touches from below, the lower
error bars of the lattice QCD data points (except for the two upper data points at T=125 MeV and 130 MeV
and lower data points after 180 MeV).} 
\label{fig:mini:fig1} %% label for entire figure
\end{figure*}

For $T=0$, the condensate $\sigma_{x0}$ = 92.4 MeV while the $\sigma_{y0}$ = 94.5 MeV, we have 
plotted the temperature variations of ($\frac{\sigma_{x}}{\sigma_{x0}}$) , ($\frac{\sigma_{y}}{\sigma_{y0}}$)
and $\Phi$ in Fig.\ref{fig:mini:fig1:a} for $m_{\sigma}=400 $ MeV and the Polyakov loop potential parameter $T_{glue}^{c}=240$ MeV.
We have shown the temperature variation of the $  \Delta_{l,s}(T)  $ in the Fig.\ref{fig:mini:fig1:b}. We obtained the values
of parameters $ T_{1}=\ 108 \ \text{MeV} \ = \ .70 \ T_{c}$ and  
$ b_{1} =\ 90 \ \text{MeV} \ = \ .58 \ T_{c} $  where $ T_{c} = \ 154 $ MeV for the temperature 
variation of c(T) by fitting the Wuppertal group lattice data for the $  \Delta_{l,s} (T) $ \cite{LQCDWB2} 
to our calculations in model PQMVT-II when $m_{\sigma}=400$ MeV and $T_{glue}^{c}=240$ MeV.
$ T_{c}$ denotes pseudocritical temperature for chiral crossover obtained from LQCD data. The solid deep red line in 
Fig.\ref{fig:mini:fig1:b} for the PQMVT-II model, shows a very good agreement of $  \Delta_{l,s} (T)  $ temperature 
variation with lattice QCD data \cite{LQCDWB2}. Table \ref{tab:pmtr} shows parameters for the temperature dependence of c(T)
for different values of $\sigma$ meson mass. The dash dotted line in green, shows temperature variation of the 
$ \Delta_{l,s} (T) $ for $T_{glue}^{c}=260$ MeV in the PQMVT-II model while dash line in deep blue  shows $ \Delta_{l,s} (T) $ 
for $T_{glue}^{c}=220$ MeV. The right most line in magenta dots denotes $ \Delta_{l,s} (T) $ temperature variation in the 
PQMVT-I model for constant c.

The dash line in green in Fig.\ref{fig:mini:fig1:a}, shows the scaled non-strange condensate temperature
variation in the PQMVT-I model for constant c and its  inflection point gives the pseudo-critical temperature as 
$T_{c}^{\chi}=\ 175.6$ MeV while the dotted line in green shows the corresponding strange condensate temperature variation.
The second peak of the strange condensate derivative $\frac{\partial\sigma_y}{\partial T}$ is rather flat and broad and it gives
$T_{s}^{\chi}=\ 261.7 \pm 2$ MeV where $\pm$ denotes a distinct change of about 0.1 percent in the numerical value of the second 
peak height.
In PQMVT-II model, for the temperature dependent c(T), the deep red solid line shows a large quantitative and a little
sharper qualitative change in the temperature variation of non-strange condensate 
when compared to the PQMVT-I scenario because the temperature variation of $\frac{\partial\sigma_x}{\partial T}$ shows a 
a little sharper and higher peak (Fig not shown in the text) which occurs quite early at $ T_{c}^{\chi}= \ 154.9 $ MeV in 
the PQMVT-II scenario. The deep red dash dotted line, shows the corresponding strange condensate temperature variation in
the PQMVT-II model, it shows large change in comparison to the PQMVT-I model for the temperature range 125-210 MeV. 
The second peak of the strange condensate gives the pseudo-critical temperature $ T_{s}^{\chi} = 265.2 \pm 2 $ MeV for the 
PQMVT-II model.

Since we have taken  $\mu=0$ in our computations, we get $\Phi$ \ = \ $\bar\Phi$ for the Polyakov loop expectation value.  
We recall that the logarithmic contribution in the improved ansatz of the polyakov loop potential \cite{Haas} in 
Eq. (\ref{pot:improved}) avoids the $\Phi$ expectation value higher than one. It represents more effective description of the gluon 
dynamics. The inflection point of the temperature variation of the Polyakov loop expectation value, shown by the green line with
filled triangles in Fig.\ref{fig:mini:fig1:a} , gives confinement-deconfinement transition pseudocritical temperature 
$T_{c}^{\Phi}=155.3 $ MeV, in the PQMVT-I model for constant c. The red line with hollow circles,show the temperature variation
of the Polyakov loop field $\Phi$ in the PQMVT-II model scenario for temperature dependent c(T) whose inflection point gives 
$T_{c}
^{\Phi}=145.8 $ MeV when $m_\sigma$=400 MeV. $T_{glue}^{c}=240$ MeV for all the cases. When $m_\sigma$=500 MeV,
$T_{c}^{\Phi}=154.3 $ MeV in PQMVT-I model while $T_{c}^{\Phi}=146.1 $ MeV in PQMVT-II scenario.
$T_{c}^{\Phi}=155.3 $ MeV in PQMVT-I model and $T_{c}^{\Phi}=146.1 $ MeV in PQMVT-II scenario for $m_\sigma$=600 MeV
as shown in Table \ref{tab:ctemp}. 

%%%%%%%%%%%%%%%%%%%%%%%%%%%%%%%%%%%%%%%%%%%%%%
\begin{figure*}[!htbp]
\subfigure[Subtracted chiral condensate $\Delta_{l,s}$ variation with temperature.]{
\label{fig:mini:fig2:a} %% label for first subfigure
\begin{minipage}[b]{0.43\linewidth}
\centering \includegraphics[width=3.15in]{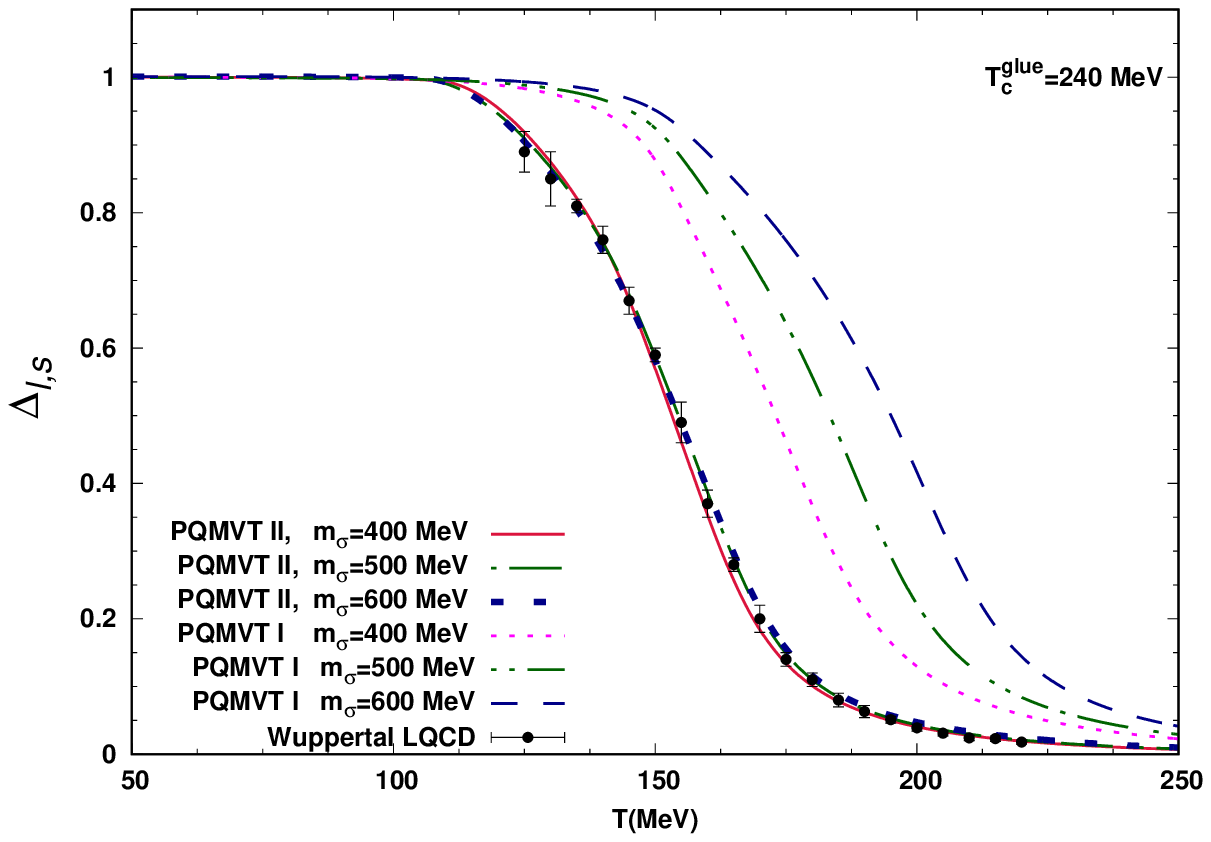}
\end{minipage}}%
\hspace{.25in}
\subfigure[Variation of c(T) with temperature.]{
\label{fig:mini:fig2:b} %% label for second subfigure
\begin{minipage}[b]{0.43\linewidth}
\centering \includegraphics[width=3.15in]{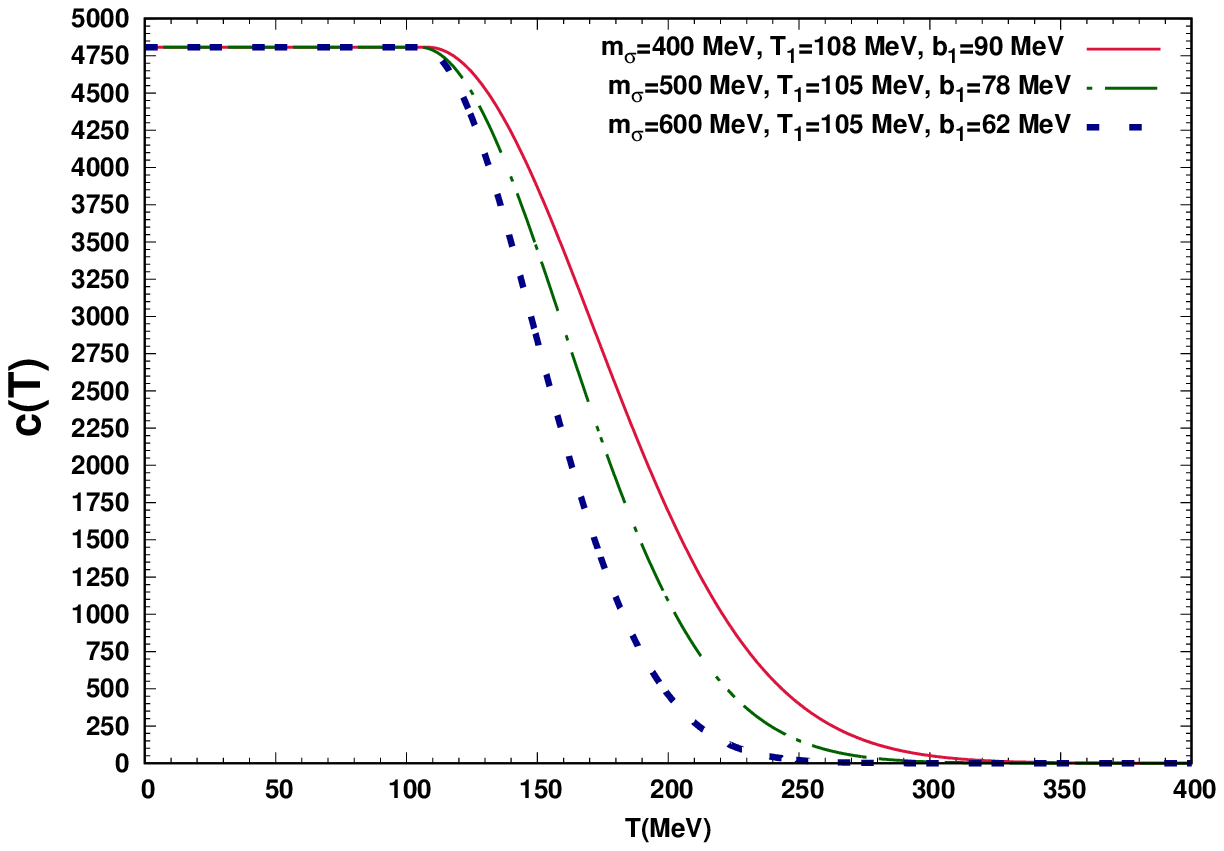}
\end{minipage}}
\caption{In Fig.\ref{fig:mini:fig2:a}, when $m_{\sigma}$=400 MeV, dotted line in magenta shows temperature variation 
of $\Delta_{ls}$ for constant KMT coupling in PQMVT I model while solid red line depicts $\Delta_{ls}$  variation for 
temperature dependent coupling c(T) 
in PQMVT II model when $T_1$=108 MeV and $b_1$=90 MeV. When $m_{\sigma}$=500 MeV,  dash double dot line in green shows
temperature variation of $\Delta_{ls}$ for constant KMT coupling in PQMVT I model while dash dot line in green depicts
$\Delta_{ls}$  variation for temperature dependent coupling c(T) in PQMVT II model when $T_1$=105 MeV and $b_1$=78 MeV.
When $m_{\sigma}$=600 MeV, long thin dashed line in blue shows temperature variation of $\Delta_{ls}$ for constant KMT 
coupling in PQMVT I model while thick dashed line in blue depicts $\Delta_{ls}$  variation for temperature dependent coupling
c(T) in PQMVT II model when $T_1$=105 MeV and $b_1$=62 MeV. $T_{glue}^{c}=240$ MeV for all the cases. The Wuppertal 
group LQCD data points within the error bars for the $\Delta_{ls}$ temperature variation coincide quite well
with the corresponding PQMVT-II model calculations ( for the temperature dependent 
coupling c(T))  of the $\Delta_{ls}$  as  shown by solid red line (when $m_{\sigma}$=400 MeV,$T_1$=108 MeV and $b_1$=90 MeV), 
dash dot line in green ( when $m_{\sigma}$=500 MeV,$T_1$=105 MeV and $b_1$=78 MeV )  and thick small dash line 
in blue ( when $m_{\sigma}$=600 MeV,$T_1$=105 MeV and $b_1$=62 MeV).
Fig.\ref{fig:mini:fig2:b}, shows variation of c(T) with temperature. When $m_{\sigma}$ = 400 MeV, solid red line shows 
temperature variation of c(T) for parameters $T_1$=108 MeV and $b_1$=90 MeV. Dash dot line in green shows temperature 
variation of c(T) for parameters $T_1$=105 MeV and $b_1$=78 MeV when $m_{\sigma}$=500 MeV. Thick blue dash line shows 
temperature variation of c(T) for parameters $T_1$=105 MeV and $b_1$=62 MeV when $m_{\sigma}$=600 MeV.}
\label{fig:mini:fig2} %% label for entire figure
\end{figure*}
%%%%%%%%%%%%%%%%%%%%%%%%%%%%%%%%%%%%%%%%%%%%%%%%%%%%%%%%%%%%%%%%%%%%%%%%%%%%%%%%%%%%%%%%%%%%%%%%%%%%%

Higher values of $m_\sigma$ makes the non strange condensate melting smoother and shifts it to quite high 
temperatures in the PQMVT-I model. The $\Delta_{l,s}(T)$ temperature variation for the constant c also shifts to quite high
temperatures as shown by dash double dot line in green for $m_\sigma$=500 MeV and long thin dash line in blue
for $m_\sigma$=600 MeV in Fig.\ref{fig:mini:fig2:a} where for comparison, we have repeated the dotted line plot 
in magenta color for $m_\sigma$=400 MeV (already given in Fig.\ref{fig:mini:fig1:b}). The pseudocritical 
temperature $T_{c}^{\chi}$, which is 175.6 MeV for $m_\sigma$=400 MeV and constant c in PQMVT-I scenario,
shifts to  $T_{c}^{\chi}$=187.9 MeV for $m_\sigma$=500 MeV and $T_{c}^{\chi}$=201.3 MeV for $m_\sigma$=600 MeV.
Here $T_{glue}^{c}=240$ MeV for all the cases. This effect of higher $m_\sigma$ on $\Delta_{l,s}(T)$ temperature variation,
is completely offset by having a sharper and steeper decrease of c(T) at lower $T_1$ and $b_1$ values in the PQMVT-II scenario
and we could find a good fit to the Wuppertal group LQCD data for $\Delta_{l,s}(T)$  also for $m_{\sigma}$=500 and 600 MeV.
 The LQCD data points within the error bars for the $\Delta_{ls}(T)$ coincide quite well with the corresponding calculations
having the temperature dependence of coupling c(T) as shown in Fig.\ref{fig:mini:fig2:a} by solid red line 
for $m_{\sigma}$=400 MeV,$T_1$=108 MeV=.70$T_{c}$ and $b_1$=90 MeV=.58$T_{c}$, dash dot line in green for $m_{\sigma}$=500 MeV,
$T_1$=105 MeV=.68$T_{c}$ and $b_1$=78 MeV=.51$T_{c}$ and thick small dash line in blue for $m_{\sigma}$=600 MeV,
$T_1$=105 MeV=.68$T_{c}$ and $b_1$=62 MeV=.40$T_{c}$. In the PQMVT-II model scenario, the pseudocritical temperature 
$T_{c}^{\chi}$=154.9 MeV for $m_\sigma$=400 MeV is quite close to the $T_{c}^{\chi}$=158.3 MeV for $m_\sigma$=500 MeV 
and $T_{c}^{\chi}$=159.0 MeV for $m_\sigma$=600 MeV. Since the temperature dependence of KMT coupling causes quite early 
and a little sharper melting of the non-strange condensate in the PQMVT-II scenario, we have plotted the temperature 
variation of c(T) in Fig.\ref{fig:mini:fig2:b}. When $m_{\sigma}$ = 400 MeV, solid red line shows 
temperature variation of c(T) for parameters $T_1$=108 MeV and $b_1$=90 MeV. The temperature variation of c(T)
becomes steeper with the parameters $T_1$=105 MeV and $b_1$=78 MeV for $m_{\sigma}$=500 MeV as shown by dash dot line in green.
In order to find a good fit to the $\Delta_{l,s}(T)$ variation with the LQCD data and offset the 
effect of higher value of $m_{\sigma}$=600 MeV, the  temperature variation of c(T) becomes
steepmost in the PQMVT-II scenario for the parameters $T_1$=105 MeV and $b_1$=62 MeV as plotted by thick blue dash line 
in Fig.\ref{fig:mini:fig2:b}.

In all the  situations, confinement-deconfinement transition occurs earlier than the chiral transition as 
$T_{c}^{\Phi}< T_{c}^{\chi} $ but this difference is least of about 9 MeV for the PQMVT-II model with 
temperature dependent c(T) when $m_{\sigma}=400$ MeV, $T_{glue}^{c}=240$  MeV while with this parameter set for
the constant c scenario in the model PQMVT-I, this difference is large 20.3 MeV. The difference $T_{c}^{\chi}-T_{c}^{\Phi}$ 
is largest of about 46 MeV for the PQMVT-I model with constant c when $m_{\sigma}=600$ MeV, $T_{glue}^{c}=240$ MeV
while with this parameter set for the temperature dependent c(T) in the PQMVT-II scenario, this difference is only of 
12.9 MeV. The larger value of sigma meson mass badly spoils the closeness of chiral crossover with the confinement-deconfinement 
transition while temperature dependence of anomaly coefficient c(T) brings the confinement-deconfinement crossover
transition closer to the chiral crossover transition.
Due to temperature
dependence of c(T), the chiral crossover transition temperature $ T_{c}^{\chi} $ gets quite reduced in the PQMVT-II model 
when compared to the constant c scenario, respectively
by 20.7 MeV (for $m_{\sigma}=400$ MeV), 33.6 MeV (for $m_{\sigma}=500$ MeV) 
and 42.3 MeV (for $m_{\sigma}=600$ MeV) while $T_{glue}^{c}$ is fixed at 240 MeV. The Physics of axial anomaly is 
microscopically interconnected with the confinement-deconfinement transition and chiral crossover transition
in the coupled equations of our model. Temperature dependence  of c(T) triggers early  confinement-deconfinement transition 
which gives rise to early chiral crossover transition .

\begin{figure*}[!htbp]
\subfigure[\ Constant KMT coupling c for $U_A(1)$ anomaly ]{
\label{fig:mini:fig3:a} %% label for first subfigure
\begin{minipage}[b]{0.45\linewidth}
\centering \includegraphics[width=3.15in]{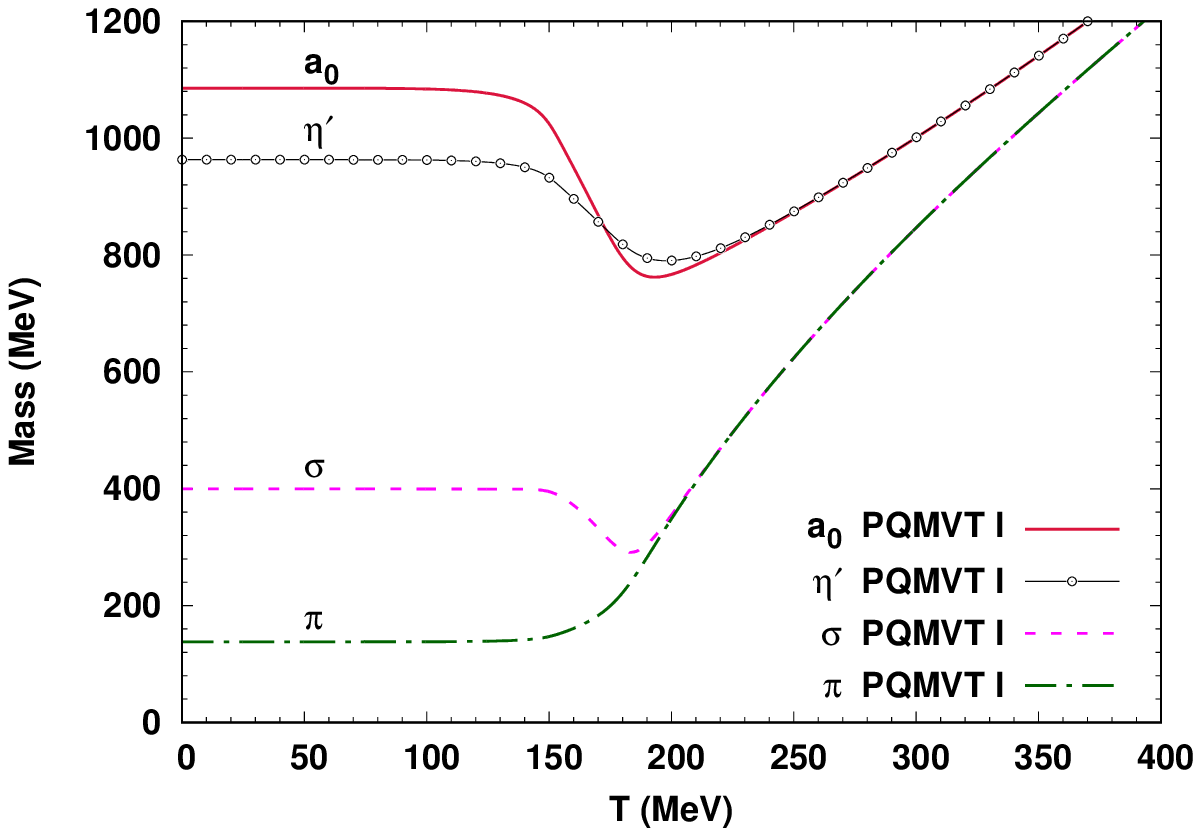}
\end{minipage}}%
\hspace{0.10in}
\subfigure[\ Temperature dependent KMT coupling c(T) for $U_A(1)$ anomaly]{
\label{fig:mini:fig3:b} %% label for second subfigure
\begin{minipage}[b]{0.45\linewidth}
\centering \includegraphics[width=3.15in]{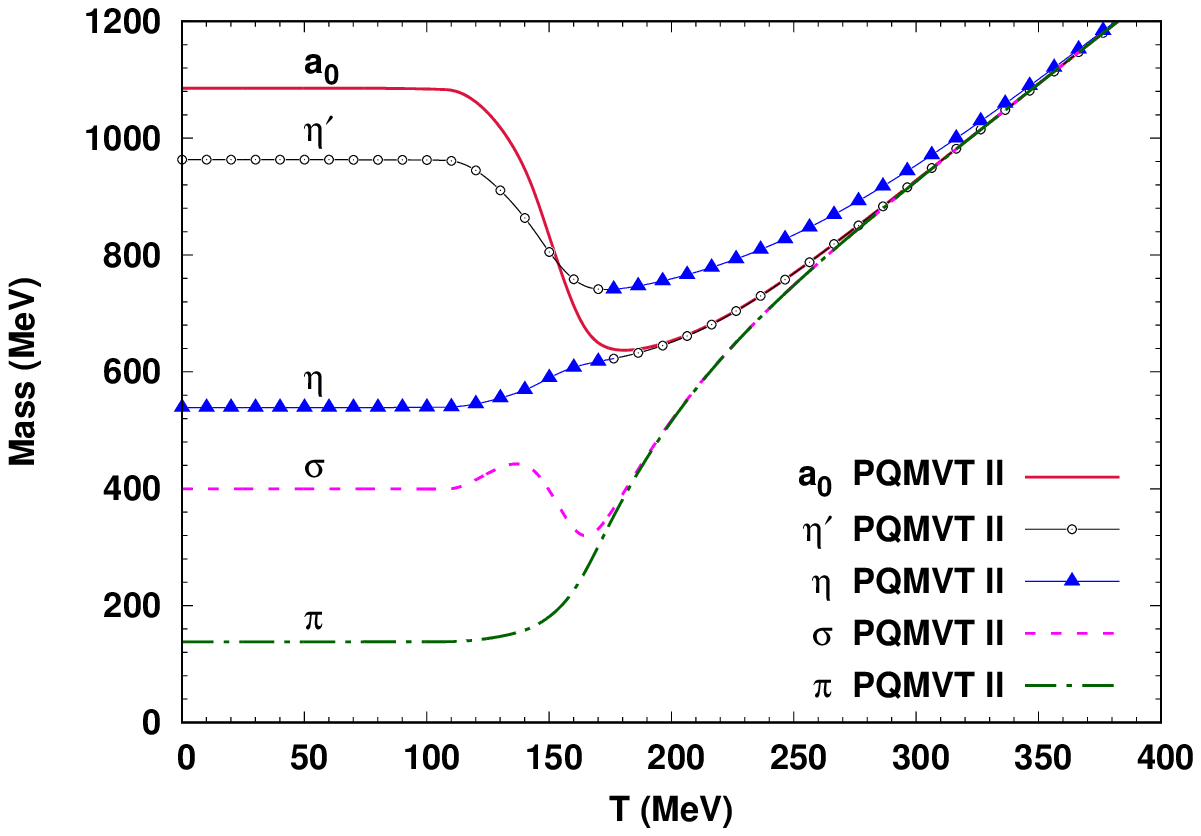}
\end{minipage}}
\caption{ Shows the mass variations for the ($\pi$, $\sigma$) and ($\eta'$, $a_{0}$) with
respect to the temperature at zero chemical potential ($\mu = 0$),$m_{\sigma}=400$ MeV and $T_{glue}^{c}=240$ MeV .
Fig.\ref{fig:mini:fig3:a} shows the results for the PQMVT-I model  and the line types
for mass variations are  labeled. Fig.\ref{fig:mini:fig3:b} shows the 
mass variations for the PQMVT-II model with labeled line types where $\eta$ mass variation has also been shown. }
\label{fig:mini:fig3} %% label for entire figure
\end{figure*}

% You have to comment about chiral transition in strange sector and confinement-deconfinement transition. 															 
%%%%%%%%%%%%%%%%%%%%%%%%%%%%%%%%%%%%%%%%%%%%%%%%%%%%%%%%%%%%%%%%%%%%%%%%%%%%%%
\subsection{ Meson Mass Variations and the $U_{A}(1)$  Restoration }
%%%%%%%%%%%%%%%%%%%%%%%%%%%%%%%%%%%%%%%%%%%%%%%%%%%%%%%%%%%%%%%%%%%%%%%%%%%%%%
 
\begin{figure*}[!htbp]
\subfigure[\ Constant KMT coupling c for $U_A(1)$ anomaly ]{
\label{fig:mini:fig4:a} %% label for first subfigure
\begin{minipage}[b]{0.45\linewidth}
\centering \includegraphics[width=3.15in]{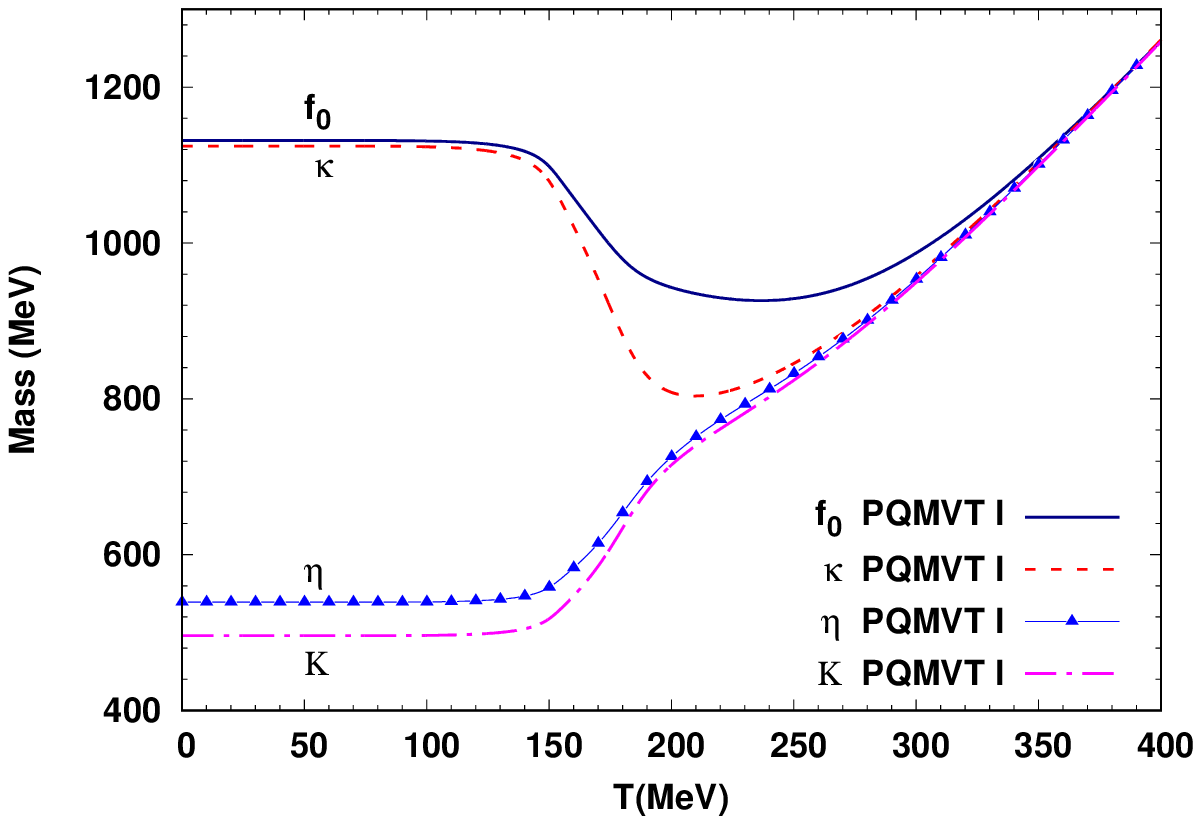}
\end{minipage}}%
\hspace{0.10in}
\subfigure[\ Temperature dependent KMT coupling c(T) for $U_A(1)$ anomaly ]{
\label{fig:mini:fig4:b} %% label for second subfigure
\begin{minipage}[b]{0.45\linewidth}
\centering \includegraphics[width=3.15in]{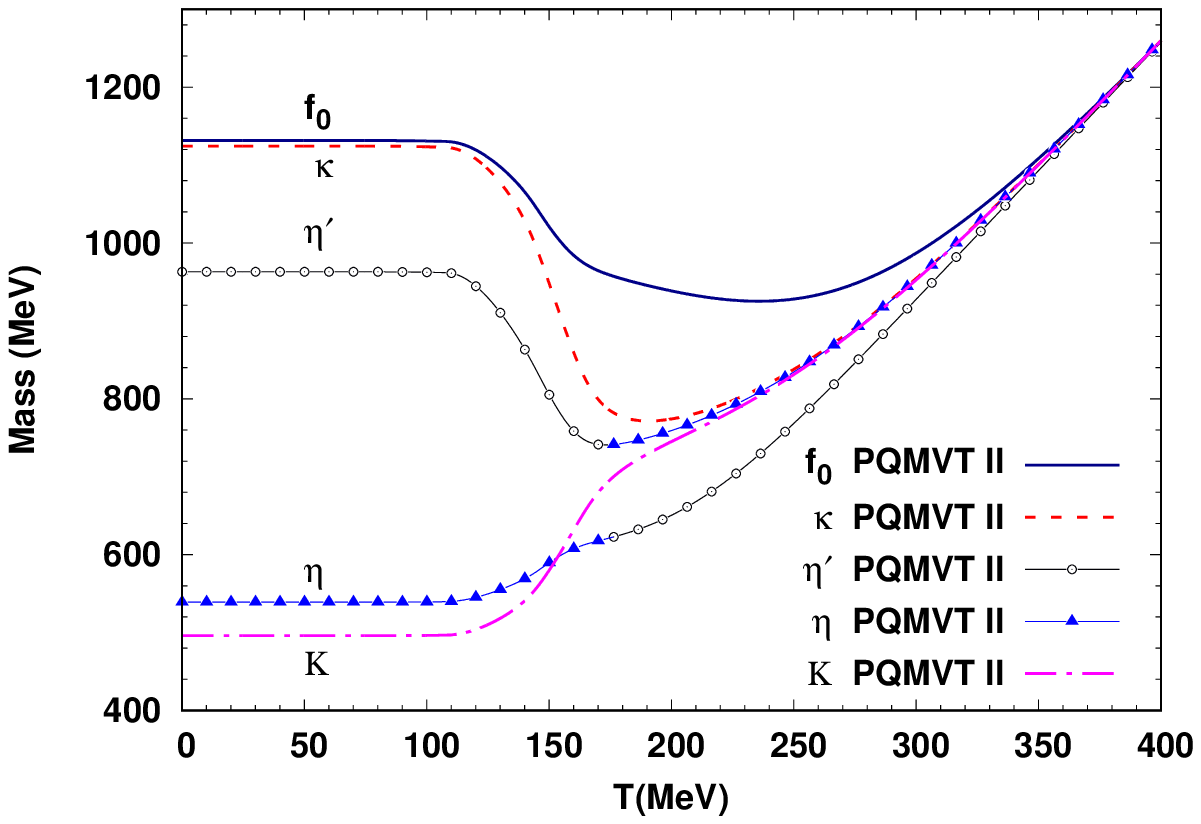}
\end{minipage}}
\caption{Shows the mass variations for the ($K$, $\kappa$) and ($\eta$, $f_{0}$) with
respect to the temperature at zero chemical potential ($\mu = 0$),$m_{\sigma}=400$ MeV and $T_{glue}^{c}=240$ MeV.
Fig.\ref{fig:mini:fig4:a} shows the results for the PQMVT-I model  and the line types
for mass variations are  labeled. Fig.\ref{fig:mini:fig4:b} shows the 
mass variations for the PQMVT-II model with labeled line types where $\eta^{\prime}$ mass variation has also been shown.}
\label{fig:mini:fig4} %% label for entire figure
\end{figure*}

Temperature variations of the meson masses ($\sigma, \pi$) and ($ \eta', a_{0} $) in the PQMVT-I model 
with constant c has been presented in the left panel (Fig.\ref{fig:mini:fig3:a}) while the corresponding 
temperature variations for the temperature dependent KMT coupling strength c(T) has been shown in the
right panel (Fig.\ref{fig:mini:fig3:b}) for the PQMVT-II model setting when $m_{\sigma}=400$ MeV, $T_{glue}^{c}=240$ MeV. 
Dash dot line in green
shows the temperature variation of $\pi$ meson, dash line in magenta colour shows the $\sigma$ mass
temperature variation while solid line with hollow circles in black, shows temperature variation of $\eta' $
meson mass and solid red line depicts temperature variation of $a_{0} $ meson mass. Showing the trend of chiral 
symmetry restoring transition in Fig.\ref{fig:mini:fig3:a} $\sigma$ degenerates with its chiral partner $\pi$ near
204 MeV  and $ a_{0} $ meson degenerates with the chiral partner $\eta'$ near 231 MeV for the PQMVT-I model. Afterwards,  
the degenerated $\sigma$, $\pi$ meson mass line, shows a poor converging trend towards the degenerated $ a_0 $, $ \eta' $ meson 
mass line, these two lines merge with each other for very large temperature of about 990 MeV$\sim 5.6 T_{c}^{\chi}$ 
(not shown in the Fig.\ref{fig:mini:fig3:a}). Persistence of $ U_{A}(1) $ symmetry in PQMVT-I upto such large temperature is 
not tenable because instanton density is supposed to vanish near $2 \ T_{c}^{\chi}$.

\begin{figure*}[!htbp]
\subfigure[\ Constant KMT coupling c for $U_A(1)$ anomaly]{
\label{fig:mini:fig5:a} %% label for first subfigure
\begin{minipage}[b]{0.45\linewidth}
\centering \includegraphics[width=3.32in]{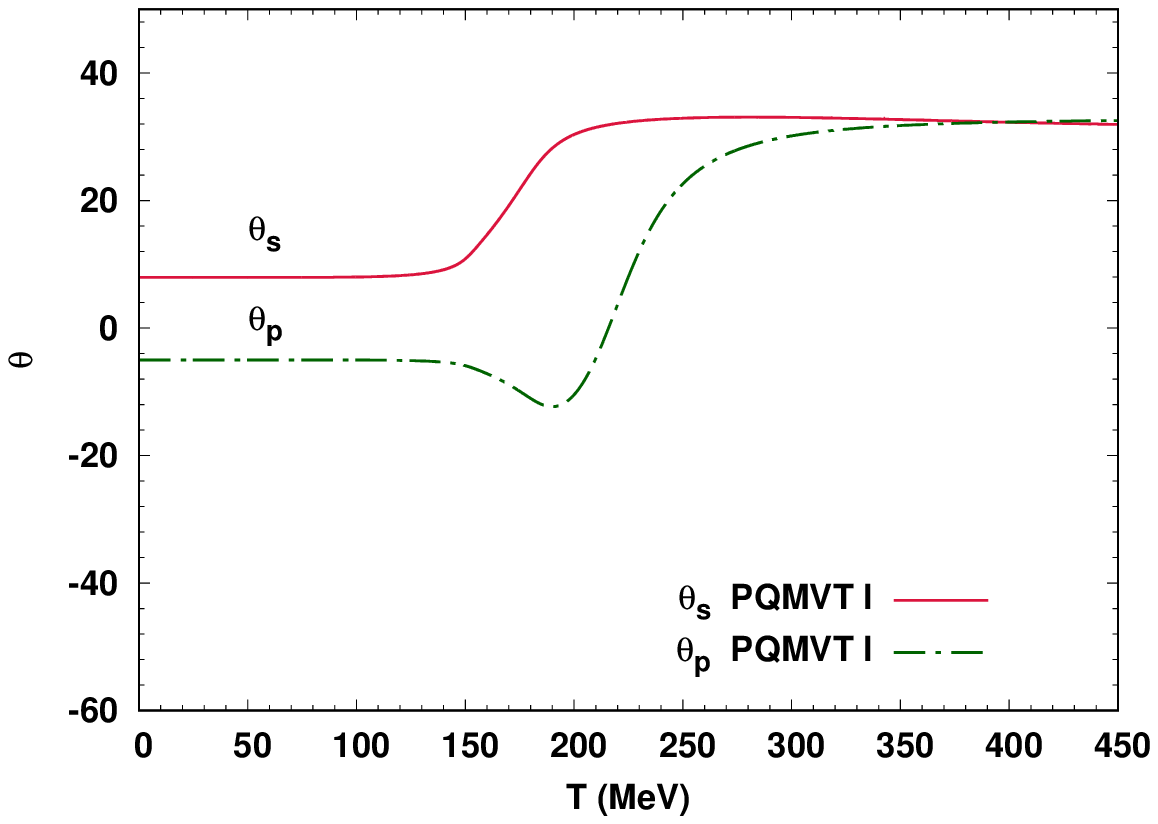}
\end{minipage}}%
\hspace{0.10in}
\subfigure[\ Temperature dependent KMT coupling c(T) for $U_A(1)$ anomaly]{
\label{fig:mini:fig5:b} %% label for second subfigure
\begin{minipage}[b]{0.45\linewidth}
\centering \includegraphics[width=3.32in]{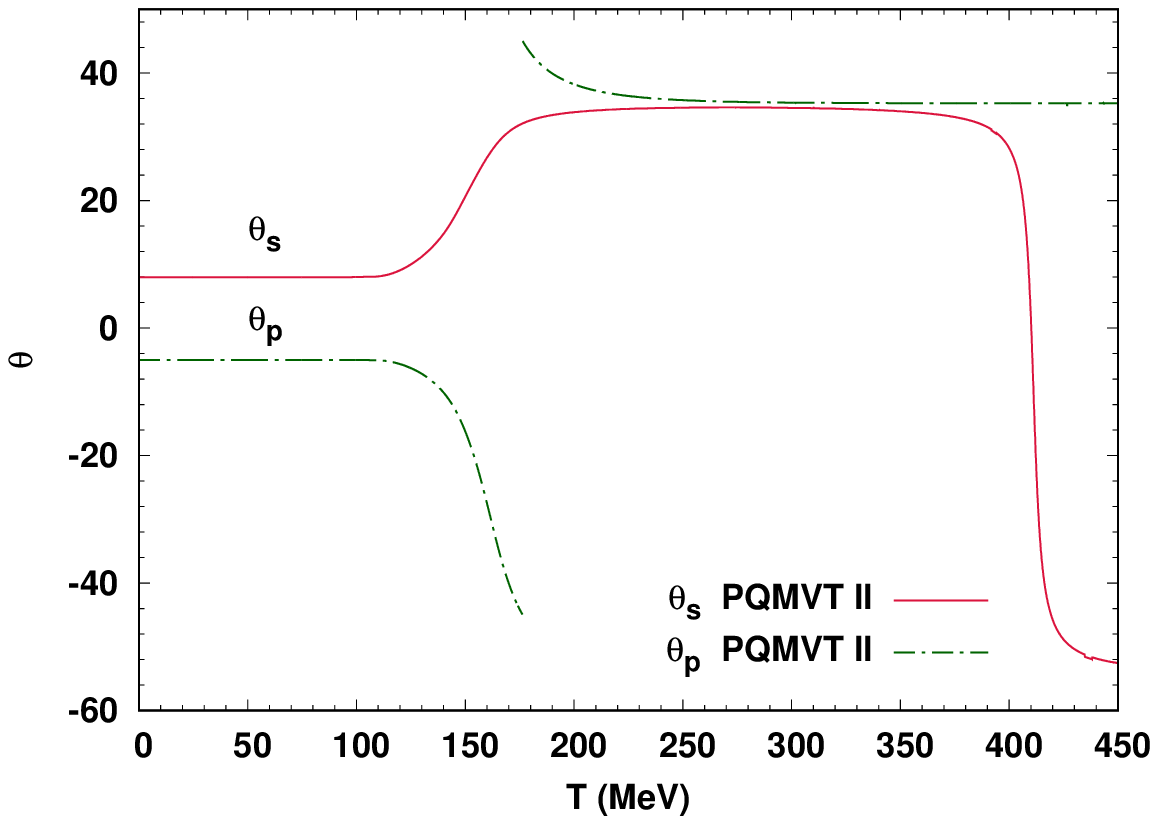}
\end{minipage}}
\caption{Shows the temperature variations of pseudoscalar mixing angle $\theta_{p}$ and scalar mixing angle $\theta_{s}$
at zero ($\mu=0$) chemical potential, $m_{\sigma}=400$ MeV and $T_{glue}^{c}=240$ MeV . Fig.\ref{fig:mini:fig5:a} shows the results for the PQMVT-I model  and the line types
for mixing angle variations are  labeled. Fig.\ref{fig:mini:fig5:b} shows the 
mixing angle variations for the PQMVT-II model with labeled line types.  }
\label{fig:mini:fig5} %% label for entire figure
\end{figure*}

\begin{figure*}[!htbp]
\subfigure[\ Constant KMT coupling c(T) for $U_A(1)$ anomaly]{
\label{fig:mini:fig6:a} %% label for first subfigure
\begin{minipage}[b]{0.45\linewidth}
\centering \includegraphics[width=3.15in]{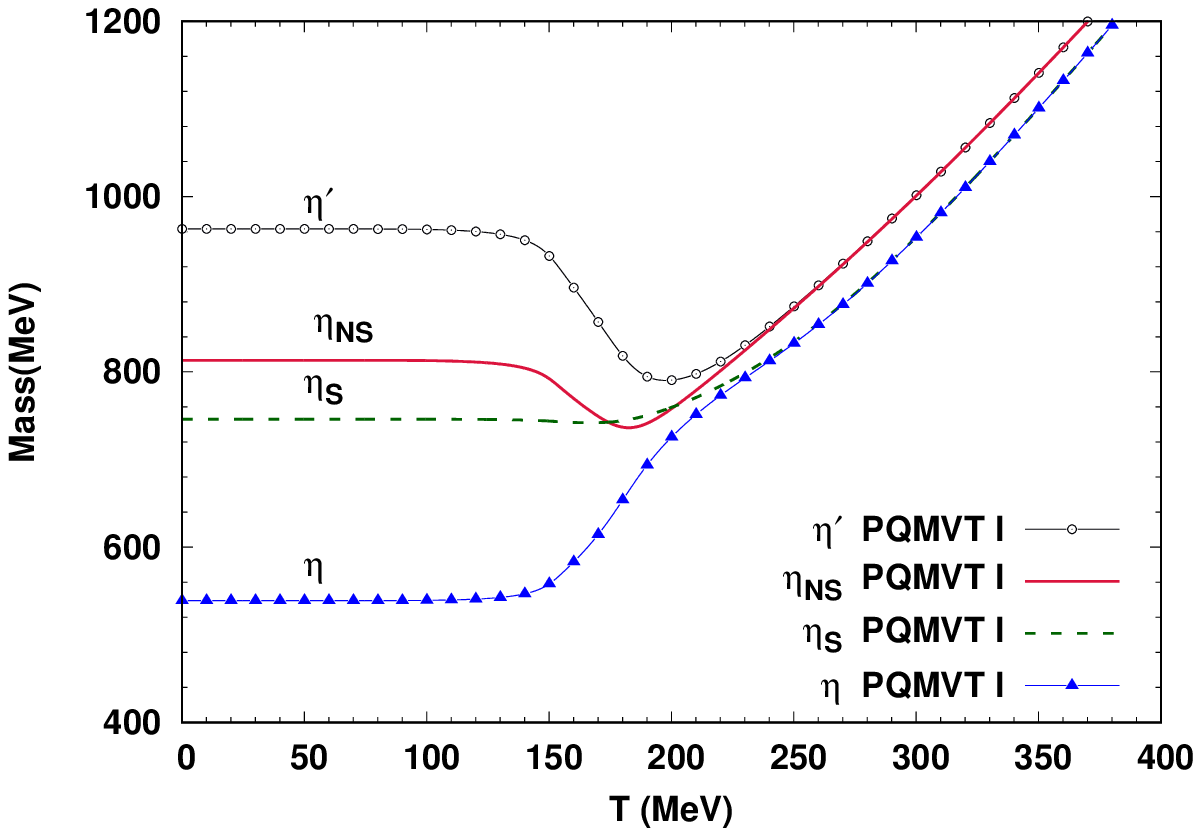}
\end{minipage}}%
\hspace{0.10in}
\subfigure[\ Temperature dependent KMT coupling c(T) for $U_A(1)$ anomaly]{
\label{fig:mini:fig6:b} %% label for second subfigure
\begin{minipage}[b]{0.45\linewidth}
\centering \includegraphics[width=3.15in]{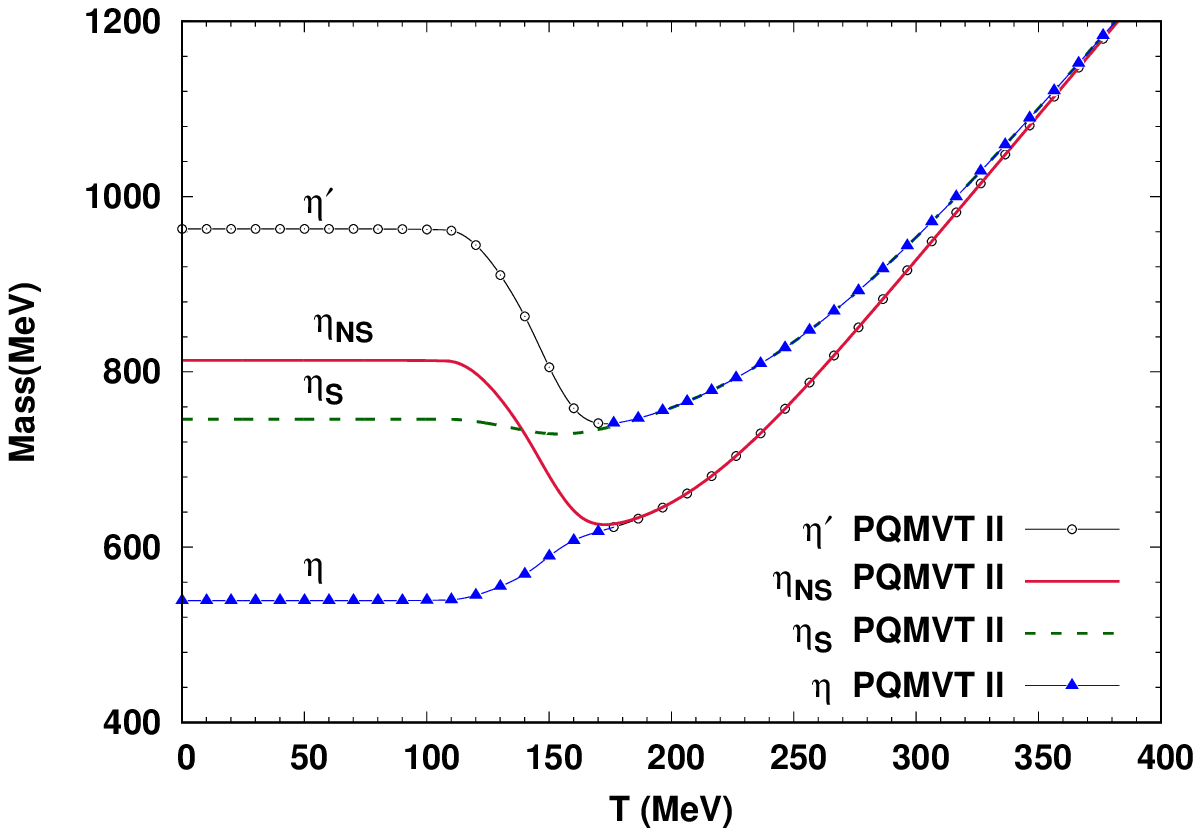}
\end{minipage}}
\caption{Shows the mass variations for the physical $\eta'$, $\eta$ and the 
non strange-strange $\eta_{NS}$, $\eta_{S}$ complex, with respect to 
temperature at zero chemical potential ($\mu = 0$), $m_{\sigma}=400$ MeV and $T_{glue}^{c}=240$ MeV.
Fig.\ref{fig:mini:fig6:a} shows the results for PQMVT-I model and line types
for mass variations are  labeled. Fig.\ref{fig:mini:fig6:b} shows the 
mass variations for the PQMVT-II model with labeled line types.}     
\label{fig:mini:fig6} %% label for entire figure
\end{figure*}

We find that due to the effect of c(T) in the PQMVT-II scenario as shown in Fig.\ref{fig:mini:fig3:b}, the $ \eta^{\prime}$ 
meson mass decreases from its vacuum value by 220 MeV near T=176 MeV after the chiral transition (at $ T_{c}^{\chi}=154.9$ MeV) temperature.
This is similar to the $ \eta^{\prime}$ in-medium mass drop of at least 200 MeV as reported by Csorgo and Vertesi in 
Ref \cite{Csorgo,Vertesi}, as an experimental signature of the effective restoration of $U_A(1)$ symmetry. 
In the PQMVT-I model for constant c case also as shown in Fig.\ref{fig:mini:fig3:a}, the $ \eta^{\prime}$ 
meson mass decreseas by 170 MeV near T=200 MeV after the chiral transition temperature (at $ T_{c}^{\chi}=175.6$ MeV). 
We have shown the temperature variation of $ \eta $ meson also in Fig.\ref{fig:mini:fig3:b} by the blue line with 
filled triangles for the temperature dependent coupling c(T), in the PQMVT-II model. We point out 
that the temperature variation in the mass of $ \eta' $ shows a discontinuous drop of about 118 MeV at
176 MeV. This happens because, the temperature variation of the pseudoscalar mixing angle presented later in 
Fig.\ref{fig:mini:fig5:b} shows a discontinuous jump from $-45^\circ$ to $+45^\circ$ at the same temperature. Corresponding 
discontinuous jump of 118 MeV in the temperature variation of the mass of $ \eta $ meson 
is seen exactly at the same temperature in Fig.\ref{fig:mini:fig3:b}. The $ \eta $ meson becomes light quark
system ($ \eta_{NS} $) at T=176 MeV and changes its identity with $ \eta' $ meson which becomes strange 
quark system ($ \eta_{S} $) as shown in Fig.\ref{fig:mini:fig6:b}.
%%%%%%%%%%%%%%%%%%%%%%%%%%%%%%%%%%%%%%%%%%%%%%%%%%%%%%%%%%%%%%%%%%%%%%%%%%%%%%%

%%%%%%%%%%%%%%%%%%%%%%%%%%%%%%%%%%%%%%%%%%%%%%%%%%%%%%%%%%%%%%%%%%%%%%%%%%%%%%%
  
  The mass of $\sigma$ meson, degenerates 
with that of $\pi$ near 189 MeV and mass of $ a_0 $ meson, degenerates with that of $ \eta $ (which has changed identity
with $ \eta' $) near 193 MeV. We note that the  $m_{\sigma}$ temperature variation in Fig.\ref{fig:mini:fig3:b},
shows an increase after T=108 MeV (equal to $T_1$ where the temperature dependence of c(T) begins) and reaches maximum 
at T=136.7 MeV registering an increase of about 42.5 MeV which is an uncommon behavior, afterwards it shows the usual 
behavior and becomes degenerate with the $m_{\pi}$ temperature variation. The origin of this pattern is consequence of 
decrease in c(T) which affects the squared masses $m_{s,00}^2$, $m_{s,88}^2$ and $m_{s,08}^2$ \cite{Trvk} in such a way 
that the expression of $m_{\sigma}^2$ on its numerical computation, first 
shows an uncommon increase and then decreases with temperature. Further,  
We emphasize here that for the temperature dependence of c(T) in the PQMVT-II scenario, 
the chiral partner of $ a_0 $ meson is $ \eta $ meson rather than 
the $ \eta^{\prime} $ meson, similar to the very recent findings of the FRG investigation under the $\text{LPA}^{\prime}$+Y (full 
effective potential with the running of Yukawa coupling and non vanishing anomalous dimensions) approximation  
in Ref. \cite{Rennecke} and the vector meson extended linear sigma model investigation in Ref. \cite{Kovacs}. This
finding differs also from the result of earlier studies in the linear sigma model \cite{Rischke:00}, QM model \cite{Schaefer:09},
PQM model \cite{gupta}, PQMVT model \cite{Trvk} as well as an FRG investigation  with the QM model under LPA 
(Local potential approximation) \cite{Mitter}. Such a change of identity from  $ \eta^{\prime} $ to $ \eta $ has also been 
found for density dependent KMT coupling $g_{D}(\rho)$ in the NJL model investigation in Ref. \cite{Costa:05}.

The temperature variations of the masses of the degenerated $\sigma$, $\pi$ meson line completely, merges with the temperature variations
of the masses of degenerated $a_{0}$, $\eta$ meson line near 275 MeV. It means that for the parameters of the PQMVT-II 
model when $m_{\sigma}=400$ MeV, $T_{glue}^{c}=240$ MeV, the $U_{A}(1)$ restoration  takes place around 275 MeV which is equal 
to 1.75 $T_{c}^{\chi}$. We point out that
mesonic thermal and quantum fluctuations investigated under FRG framework strengthen the $U_{A}(1)$ anomaly near $T_{c}^{\chi}$
\cite{Fejos, Fejos:16, Rennecke}. In Ref. \cite{Jiang}, when fluctuations are included in the calculation of topological 
susceptibility, fluctuation induced part of the  topological susceptibility remains remarkably large 
(upto T=250 MeV) in the chiral symmetry restored phase. It shows noticeable survival of $U_{A}(1)$ anomaly upto 250 MeV.
In the EPNJL model studies by Ishii et. al. \cite{Ishii}, the temperature variations of  the screening masses of $a_{0}$ meson 
and $\pi$ mesons matches with the lattice QCD data from Ref. \cite{ChengLQCD} and masses become degenerate near T=180 MeV
where they find $T_{c}^{\chi}$=163 MeV , it means $U_{A}(1)$ symmetry is restored near $T=1.1 \ T_{c}^{\chi}$. However, 
recent lattice simulations \cite{Sharma,Dick}, do not observe $U_{A}(1)$ symmetry restoration even at $T=1.5 \ T_{c}^{\chi}$
using highly improved staggered fermions. 

%%%%%%%%%%%%%%%%%%%%%%%%%%%%%%%%%%%%%%%%%%%%%%%%%%%%%%%%%%%%%%%%%%%%%%%%%%%%%%%
  
%%%%%%%%%%%%%%%%%%%%%%%%%%%%%%%%%%%%%%%%%%%%%%%%%%%%%%%%%%%%%%%%%%%%%%%%%%%%%%

    Fig.\ref{fig:mini:fig4:a} in the left panel, shows the temperature variations of masses of 
($\eta$,\ $f_0$) and ($K$, \ $\kappa$) for the constant c case in the PQMVT model while the right panel 
 Fig.\ref{fig:mini:fig4:b} presents the corresponding temperature variations of masses in the
PQMVT-II model for the temperature dependent coupling c(T). Dash dot line in magenta
shows the temperature variation of $K$ meson, solid blue color line with filled triangles, shows the $\eta$ mass
temperature variation while dash line in red , presents temperature variation of $\kappa $
meson mass and solid deep blue line shows temperature variation of $f_{0} $ meson mass.  Mass variation lines of
$K$, $\kappa$ and $\eta$ merge with each other near 339 MeV 
while $f_0$ meson mass variation almost merges with these degenerated three lines near 375 MeV in the PQMVT-I model.
In contrast, in Fig.\ref{fig:mini:fig4:b} in the right panel, for the temperature dependent c(T) in the PQMVT-II model,
mass variation lines of $K$, $\kappa$ and $\eta^{\prime}$ ($\eta$ becomes $\eta^{\prime}$ as strange quark system $\eta_{S}$
at T=176 MeV after showing jump) merge early with each other near 
270 MeV while $f_0$ meson mass variation merges with these degenerated three lines near 367 MeV. 
Here $\eta^{\prime}$ rather than $\eta$ becomes chiral partner of $f_0$ meson similar to the
FRG study result in Ref. \cite{Rennecke} under the $\text{LPA}^{\prime}$+Y approximation.   
We remind that most of the 
changes in the strange condensate,takes place in the temperature range 125 MeV to 210 MeV due to temperature dependent 
parameterizations of KMT coupling c(T) in the model PQMVT-II and the melting trend of the strange condensate variation after 
210 MeV in Fig.\ref{fig:mini:fig1:a} matches almost exactly with the corresponding strange condensate variation in the PQMVT-I 
model with constant c. Due to this, $f_0$ meson mass variation in the PQMVT-I model in Fig.\ref{fig:mini:fig4:a} looks similar 
to the temperature variation of $f_0$ mass in the Fig.\ref{fig:mini:fig4:b} in PQMVT-II model.

%%%%%%%%%%%%%%%%%%%%%%%%%%%%%%%%%%%%%%%%%%%%%%%%%%%%%%%%%%%%%%%%%%%%%%%%%%%%%

%%%%%%%%%%%%%%%%%%%%%%%%%%%%%%%%%%%%%%%%%%%%%%%%%%%%%%%%%%%%%%%%%%%%%%%%%%%%%%
%%%%%%%%%%%%%%%%%%%%%%%%%%%%%%
%%%%%%%%%%%%%%%%%%%%%%%%%%%%%%%%%%%%%%%%%%%%%%%%
\subsection{Meson Mixing Angle Variations}
%%%%%%%%%%%%%%%%%%%%%%%%%%%%%%%%%%%%%%%%%%%%%%%%%%%%%%%%%%%%%%%%%%%%%%%%%%%%%%

%%%%%%%%%%%%%%%%%%%%%%%%%%%%%%%%%%%%%%%%%%%%%%%%%%%%%%%%%%%%%%%%%%%%%%%%%%%%%%% Figure 6(a) and 6(b)
%%%%%%%%%%%%%%%%%%%%%%%%%%%%%%%%%%%%%%%%%%%%%%%%%%%%%%%%%%%%%%%%%%%%%%%%%%%%%%

We will now be exploring the nature of the scalar $\theta_s$ and pseudo scalar 
$\theta_p$ mixing angles temperature variations.
In Fig.\ref{fig:mini:fig5:a} in the left panel, the dash dotted line in green shows the pseudoscalar 
mixing angle $\theta_p$ temperature variations while the solid line in deep red shows
scalar mixing angle $\theta_s$ temperature variations for the PQMVT-I model with constant coupling c. 
In Fig.\ref{fig:mini:fig5:b} in the right panel, the same line types respectively, show the 
$\theta_p$ and $\theta_s$ temperature variations for the PQMVT-II model with temperature dependent 
KMT coupling c(T). As in Refs. \cite{Schaefer:09,gupta,Trvk}, in the low temperature chiral symmetry broken 
phase , the nonstrange and strange quark mixing is strong and one gets almost constant 
pseudoscalar mixing angle $ \theta_p=-5^{\circ} $ in both the Fig.\ref{fig:mini:fig5:a} and 
Fig.\ref{fig:mini:fig5:b} . As temperature increases after 150 MeV, the $\theta_p$ variation in the PQMVT-I model
develops a dip around pseudocritical temperature $T_{c}^{\chi}=175.6$ MeV and then smoothly starts 
approaching the ideal mixing angle $\theta_p\to \arctan{\frac{1}{\sqrt{2}}}\sim 35^{\circ}$ which gets fully achieved 
for temperatures around 350 MeV in the high temperature chiral symmetry restored phase. 

\begin{figure*}[!htbp]
\subfigure[\ Constant KMT coupling c(T) for $U_A(1)$ anomaly ]{
\label{fig:mini:fig7:a} %% label for first subfigure
\begin{minipage}[b]{0.45\linewidth}
\centering \includegraphics[width=3.15in]{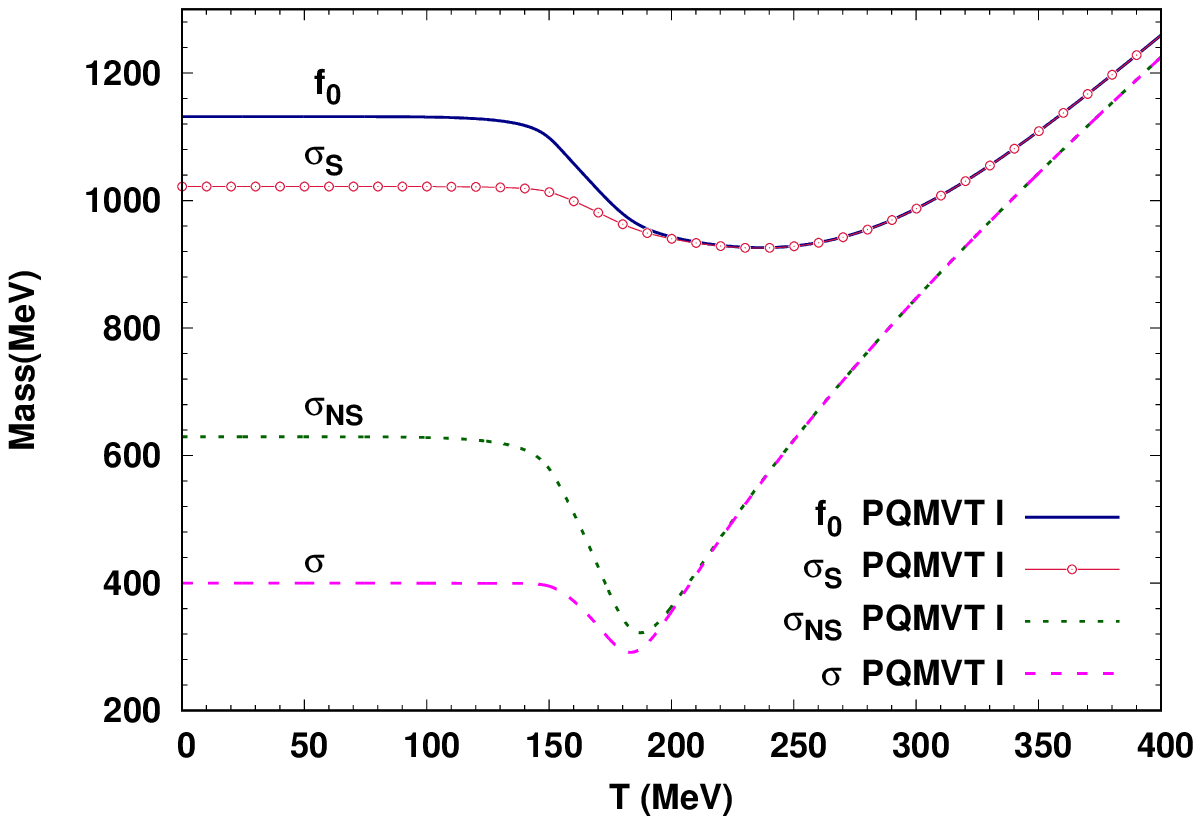}
\end{minipage}}
\hspace{0.10in}
\subfigure[\ Temperature dependent KMT coupling c(T) for $U_A(1)$ anomaly ]{
\label{fig:mini:fig7:b} %% label for second subfigure
\begin{minipage}[b]{0.45\linewidth}
\centering \includegraphics[width=3.15in]{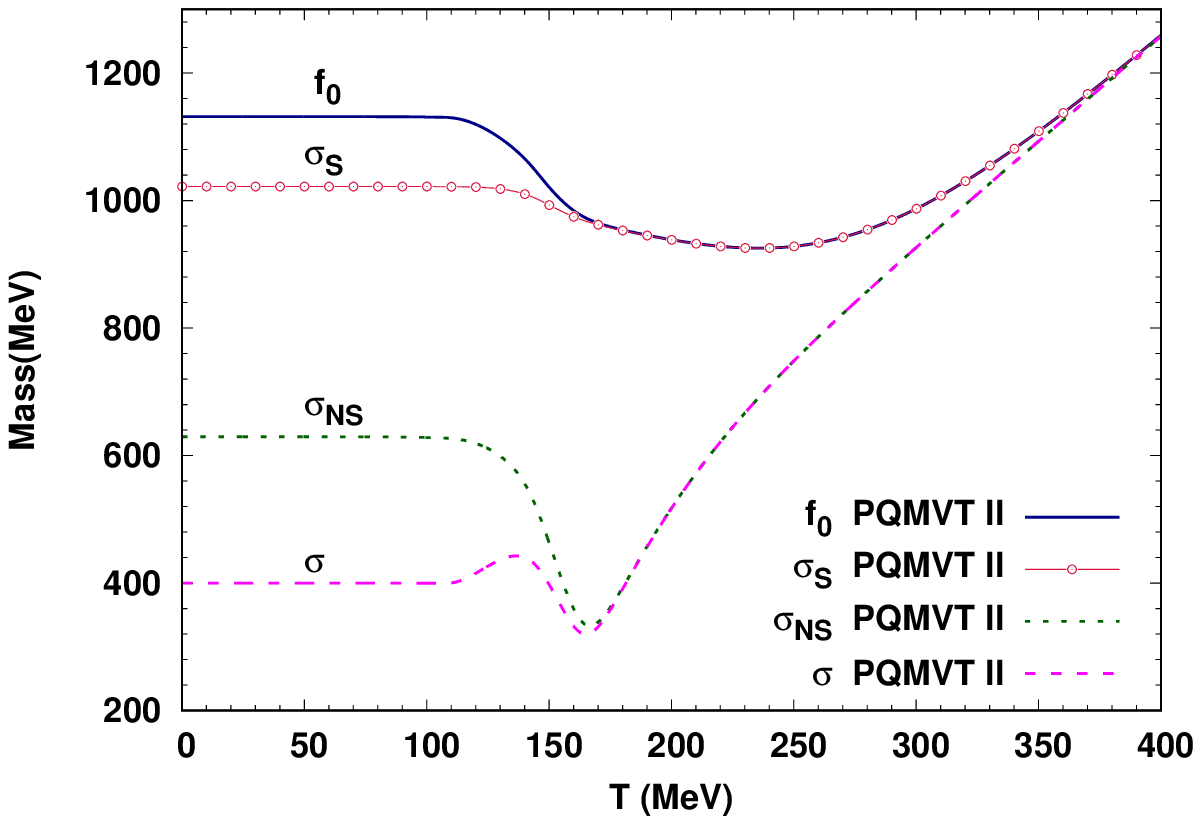}
\end{minipage}}
\caption{ Shows the mass variations for the  ($\sigma$, $\sigma_{NS}$) and for the 
 ($f_{0}$, $\sigma_{S}$) , with respect to 
temperature at zero chemical potential ($\mu = 0$), $m_{\sigma}=400$ MeV and $T_{glue}^{c}=240$ MeV.
Fig.\ref{fig:mini:fig7:a} shows the results for PQMVT-I model and line types
for mass variations are  labeled. Fig.\ref{fig:mini:fig7:b} shows the 
mass variations for the PQMVT-II model with labeled line types. }
\label{fig:mini:fig7} %% label for entire figure
\end{figure*}

The ($ \eta$,$\eta^{\prime}$) 
basis of mesons is related to the flavour nonstrange-strange ($ \eta_{NS}$,$\eta_{S}$) basis through the mixing angle 
$\phi_p $ where  $\phi_{p}=\theta_{p}+\text{arctan} \sqrt{2}\sim \theta_{p}+54.74^{\circ} $ as explained in the appendix 
C of Ref. \cite{Schaefer:09}. Thus when $\theta_p \sim 35^{\circ}$, the $\phi_p$ becomes $90^{\circ} $. Due to this
the $\eta$ and $\eta'$ mesons, respectively become a pure strange $\eta_S$ and non strange light $\eta_{NS}$ quark system
in PQMVT-I. Mass formula for $m_{\eta_{NS}}$ and $m_{\eta_S}$ are given in the Table \ref{tab:mass}.
The $m_{\eta'}$ variation  shown by the black line with hollow circles merges with  $m_{\eta_{NS}}$ variation 
depicted by the deep red line near 236 MeV in PQMVT-I model in Fig.\ref{fig:mini:fig6:a} while the  $m_{\eta}$ variation 
shown by the solid line in blue color with filled triangles merges with the $m_{\eta_S}$ variation shown by the dash line 
in green near the same temperature and further these two degenerated lines merge with each other near very high temperature of 
about 810 MeV (not shown in the Fig.\ref{fig:mini:fig6:a}).

In the PQMVT-II model,
in complete contrast, in Fig.\ref{fig:mini:fig5:b}, the  $\theta_p$ variation starts a sharp dip after 120 MeV due 
to the effect of the temperature dependent coupling $c(T)$ and at 176 MeV, the pseudoscalar mixing angle $\theta_p$
jumps from $-45^{\circ} $ to  $+45^{\circ} $. $\theta_p$ is calculated from tan$2\theta_p$ which remains positive
in first and third quadrant. When $\theta_p$ jumps from $-45^{\circ} $ to  $+45^{\circ} $, the  
observed temperature variation of $\theta_p$ afterwards from $+45^{\circ}$ to ideal mixing  $ 35.26^{\circ}$ in 
first quadrant actually results due to the variation of $\theta_p$ from $-45^{\circ}$ to anti-ideal mixing  
$-54.74^{\circ}$ in the fourth quadrant. In consequence $\phi_p $ changes from  $49.74^{\circ}$ near T=0 MeV and 
$9.74^{\circ}$ at T=176 MeV  to  $0^{\circ}$ near T=250 MeV. 
Due to this $\eta^{\prime}$ temperature variation of the mass, shows a drop of 118 MeV at T= 176 MeV 
in Fig.\ref{fig:mini:fig6:b} and changes its identity with $\eta$ meson mass variation which shows a jump of 118 MeV 
at the same temperature. $m_{\eta'}$ temperature variation degenerates with the strange quark system $m_{\eta_{S}}$ 
temperature variation after change of its identity at T= 176 MeV while $m_{\eta}$ 
temperature variation  degenerates with  $m_{\eta_{NS}}$ light quark system temperature variation.
The anti-ideal mixing when $\phi_p \to 0 $ is achieved near 250 MeV in PQMVT-II scenario.

We remind ourselves again that for the temperature dependence of c(T) in the PQMVT-II scenario, 
the behavior of the pseudoscalar mixing angle $\theta_p$ or $\phi_p$ temperature variation is similar to their corresponding 
behavior in recent findings of the FRG investigation under the $\text{LPA}^{\prime}$+Y  approximation  
in Ref. \cite{Rennecke} and the vector meson extended linear sigma model investigation in Ref. \cite{Kovacs}. This
finding differs also from the result of earlier studies in the linear sigma model \cite{Rischke:00}, QM model \cite{Schaefer:09},
PQM model \cite{gupta}, PQMVT model \cite{Trvk} as well as an FRG investigation  with the QM model under LPA 
(Local potential approximation) \cite{Mitter}. Such a jump for pseudoscalar mixing angle $\theta_p$ from $-45^{\circ} $ to 
 $+45^{\circ} $, has also been reported for density dependent KMT coupling $g_{D}(\rho)$ in the NJL model investigation in 
Ref. \cite{Costa:05}. In contrast to Fig.\ref{fig:mini:fig6:a} of PQMVT-I model, the $m_{\eta'}$ and $m_{\eta_{S}}$ 
degenerated lines further become degenerate with the $m_{\eta}$ and $m_{\eta_{NS}}$ degenerated lines i.e. 
all the four lines merge with each other near 375 MeV in Fig.\ref{fig:mini:fig6:b} for the PQMVT-II model. 
%starts approaching ideal mixing angle $ 35^{\circ}$ in 
%the high temperature chiral symmetry restored phase.     
% The effect of this 
%jump in $\theta_p$ is seen in the fall in the mass of $\eta^{'}$ meson in 
%Fig.\ref{fig:mini:fig3:b} and the corresponding jump in the mass of $\eta$ meson exactly by the same amount of 118 MeV 
%in Fig.\ref{fig:mini:fig4:b}. It means after this jump in $\theta_p$, $\eta^{'}$ and 
%$\eta$ mesons change their identity .   

%This trend indicates that $U_1(A)$ approximate restoration
% happens around 300 MeV and gets
%completed around 340 MeV
%%%%%%%%%%%%%%%%%%%%%%%%%%%%%%%%%%%%%%%%%%%%%%%%%%%%%%%%%%%%%%%%%%%%%%%%%%%%%%
%% Figure 5(a) and 5(b)
%%%%%%%%%%%%%%%%%%%%%%%%%%%%%%%%%%%%%%%%%%%%%%%%%%%%%%%%%%%%%%%%%%%%%%%%%%%%%%
 
The scalar mixing angle $\theta_s$ in vacuum ($T=0$) is about $8^{\circ}$ in both the cases of PQMVT-I model 
and PQMVT-II model in Fig.\ref{fig:mini:fig5:a} and Fig.\ref{fig:mini:fig5:b} respectively. 
The $ \theta_s $ starts growing towards its ideal value for temperature about 25 MeV higher than the 
$T_{c}^{\chi}=175.6$ MeV, in the PQMVT-I model in Fig.\ref{fig:mini:fig5:a}  and  approaches the ideal
mixing angle very smoothly in the higher temperature chirally symmetric phase. 
The effect of temperature dependent coupling c(T) modifies the temperature variation of
scalar mixing angle $\theta_s$ in the PQMVT-II model. Here, the $ \theta_s $ starts approaching 
its ideal value for temperature about 20 MeV higher than the 
$T_{c}^{\chi}=154.9$ MeV in Fig.\ref{fig:mini:fig5:b} and for higher temperatures in the chirally symmetric 
phase, the scalar mixing angle first achieves its ideal value $35^{\circ}$ and then drops down to 
$\theta_{s} \sim -53^{\circ})$ for about 450 MeV temperature. This pattern is already reported and discussed in 
Ref.\cite{Schaefer:09,gupta} for the Quark Meson (QM) model and Polyakov Quark Meson (PQM) model .

    In Fig.\ref{fig:mini:fig7:a} for the PQMVT-I model, we notice that the $f_0$ meson mass temperature
variation shown by the solid deep blue line degenerates with the pure strange quark system $\sigma_S$ mass temperature
variation shown by the solid red line with hollow circle near 196 MeV while the $\sigma$ meson mass temperature variation shown 
by the dash line in magenta becomes identical with the pure non-strange
quark system $\sigma_{NS}$ mass variation shown by the green dots line near the
temperature 209 MeV . Further these two
degenerated lines show a mild mass convergence trend and merge with each other near very high temperature of 720 MeV 
(not shown in the Fig.\ref{fig:mini:fig7:a}) . For temperature dependent coupling c(T) in the PQMVT-II model, in 
Fig.\ref{fig:mini:fig7:b}, we find that the $f_0$ meson mass temperature variation  degenerates with the 
pure strange quark system $\sigma_S$ mass temperature variation near 167 MeV and the $\sigma$ meson mass temperature variation 
merges with the pure non-strange quark system $\sigma_{NS}$ mass variation near the  
temperature 175 MeV. In quite contrast to PQMVT-I model, these two
lines show a strong mass convergence trend and merge with each other near T=385 MeV.

%%%%%%%%%%%%%%%%%%%%%%%%%%%%%%%%%%%%%%%%%%%%%%%%%%%%%%%%%%%%%%%%%%%%%%%%%%%%%%
\section{Summary and Conclusion}
\label{sec:smry}
In the present  work, we have explored, how the temperature dependence of KMT coupling strength c(T)
in the PQMVT-II model which we call as PQMVT-I model for constant coupling c , modifies the interplay of chiral symmetry and 
the $U_{A}(1)$ symmetry restoration. 
We computed the subtracted chiral order parameter $\Delta_{l,s} (T) $ temperature variation in the PQMVT-II scenario
and fitted it to the Wuppertal group LQCD data from which they obtained the pseudo-critical temperature for chiral transition as
$ T_{c}=157 (3)$ MeV \cite{LQCDWB2}. Temperature dependence of KMT coupling causes quite early and a little sharper melting of 
the non-strange condensate in the PQMVT-II scenario while larger values of $m_\sigma$ makes this melting smoother and shifts it
to quite high temperatures in the PQMVT-I model. 
% When, $m_{\sigma}=400$ MeV and Polyakov loop potentail parameter $T_{glue}^{c}=240$ MeV,
%the $ T_{c}^{\chi}=175.6$ MeV obtained in the PQMVT-I calculation, gets reduced by 20.7 MeV to become $ T_{c}^{\chi}=154.9$ 
%MeV because of the temperature dependence of c(T) in the PQMVT-II scenario. When, $m_{\sigma}=600$ MeV and $T_{glue}^{c}=240$ MeV,
%the PQMVT-I model $ T_{c}^{\chi}=198.1$ MeV becomes $ T_{c}^{\chi}=159.1$ MeV in the PQMVT-II scenario which is a large reduction of 
%39 MeV. 
In comparison to PQMVT-I model, the nonstrange condensate and strange condensate temperature variations
for the PQMVT-II scenario, are found to have the largest difference in the temperature range 125-210 MeV. 
In all the  situations, confinement-deconfinement transition occurs earlier than the chiral transition as 
$T_{c}^{\Phi}< T_{c}^{\chi} $ but this difference is least of about 9 MeV for the PQMVT-II model with 
temperature dependent c(T) when $m_{\sigma}=400$ MeV, $T_{glue}^{c}=240$  MeV while with this parameter set for
the constant c scenario in the model PQMVT-I, this difference is large 20.3 MeV.
The larger value of sigma meson mass badly spoils the closeness of chiral crossover with the confinement-deconfinement 
transition while temperature dependence of anomaly coefficient c(T) brings the confinement-deconfinement crossover
transition closer to the chiral crossover transition.

In the constant c PQMVT-I scenario,  
the degenerated $\sigma$, $\pi$ meson mass line, shows a poor converging trend towards the degenerated $ a_0 $, $ \eta' $ meson 
masses line, these two lines merge with each other for very large temperature of about 990 MeV$\sim 5.6 T_{c}^{\chi}$.
Persistence of $ U_{A}(1) $ symmetry to such large temperature is not tenable because instanton density is most likely to approach zero near 
$2 \ T_{c}^{\chi}$. We notice that due to the effect of c(T) in the PQMVT-II scenario, the $ \eta^{\prime}$ 
meson mass decreases from it vacuum value by 220 MeV near T=176 MeV after the chiral transition (at $ T_{c}^{\chi}=154.9$ MeV) temperature.
This is similar to the $ \eta^{\prime}$ in-medium mass drop of at least 200 MeV as reported by Csorgo and Vertesi in 
Ref \cite{Csorgo,Vertesi}, as an experimental signature of the effective restoration of $U_A(1)$ symmetry.
For temperature dependence of the KMT coupling, mass variations of the degenerated $\sigma$, $\pi$ mesons merges completely 
with the temperature variations of the masses of degenerated $a_{0}$, $\eta$ mesons ( note here that chiral 
partner of $ a_0 $  is $ \eta $  rather than the $ \eta^{\prime} $)   near 275 MeV. It means that for 
the parameters of the PQMVT-II model when $m_{\sigma}=400$ MeV, the $U_{A}(1)$ restoration  takes place around 275 MeV which is equal to 
1.75 $T_{c}^{\chi}$. We note that mesonic thermal and quantum fluctuations investigated under FRG framework strengthen 
the $U_{A}(1)$ anomaly near $T_{c}^{\chi}$ \cite{Fejos, Fejos:16, Rennecke}. In the EPNJL model studies by Ishii et. al. 
\cite{Ishii}, the temperature variations of  the screening masses of $a_{0}$ meson and $\pi$ mesons matches with the 
lattice QCD data from Ref. \cite{ChengLQCD} and masses become degenerate near T=180 MeV
where they find $T_{c}^{\chi}$=163 MeV , it means $U_{A}(1)$ symmetry is restored near $T=1.1 \ T_{c}^{\chi}$. However, 
recent lattice simulations \cite{Sharma,Dick}, do not observe $U_{A}(1)$ symmetry restoration even at $T=1.5 \ T_{c}^{\chi}$
using highly improved staggered fermions.    
In the PQMVT-II model, mass variation lines of $K$, $\kappa$ and $\eta^{\prime}$ ($\eta$ becomes $\eta^{\prime}$ as 
strange quark system $\eta_{S}$ at T=176 MeV) merge early with each other near 
270 MeV while $f_0$ meson mass variation merges with these degenerated three lines near 367 MeV.
Here $\eta^{\prime}$ rather than $\eta$ becomes chiral partner of $f_0$ meson.
  
 In the PQMVT-II model, due to the temperature dependent coupling $c(T)$, the pseudoscalar mixing angle 
$\theta_p$ jumps from $-45^{\circ} $ to  $+45^{\circ} $ at 176 MeV, afterwards the $\theta_p$ temperature variation  
from $+45^{\circ}$ to ideal mixing  $ 35.26^{\circ}$ in first quadrant actually results due to the variation of 
$\theta_p$ from $-45^{\circ}$ to anti-ideal mixing  $-54.74^{\circ}$ in the fourth quadrant, since $\theta_p$ is 
calculated from tan$ \ 2\theta_p$ which is positive in first and third quadrant. 
Due to this $\eta^{\prime}$ temperature variation of the mass, drops by 118 MeV at T= 176 MeV 
and changes its 
identity with $\eta$ meson mass variation which shows a jump of 118 MeV 
at the same temperature. The $ \eta $ meson becomes light quark system ($ \eta_{NS} $) at T=176 MeV and changes 
its identity with $ \eta' $ meson which becomes strange quark system ($ \eta_{S} $).
These results are similar to the recent findings of the FRG investigation under the $\text{LPA}^{\prime}$+Y  approximation  
in Ref. \cite{Rennecke} and the vector meson extended linear sigma model investigation in Ref. \cite{Kovacs} and
differ from the results of earlier studies in the linear sigma model \cite{Rischke:00}, QM model \cite{Schaefer:09}, PQM model
\cite{gupta}, PQMVT model \cite{Trvk} as well as an FRG investigation  with the QM model under LPA (Local potential approximation)
\cite{Mitter}. Such a change of identity from  $ \eta^{\prime} $ to $ \eta $ has also been reported for density dependent
KMT coupling $g_{D}(\rho)$ in the NJL model investigation in Ref. \cite{Costa:05}. 

In PQMVT-I, $\eta$ becomes $\eta_S $ and 
$\eta'$ becomes $\eta_{NS}$ due to ideal mixing, these two degenerated lines merge with each other near very high temperature of 
about 810 MeV. In PQMVT-II model due to anti-ideal mixing, the $m_{\eta'}$ and $m_{\eta_{S}}$ 
degenerated lines further merge with the $m_{\eta}$ and $m_{\eta_{NS}}$ degenerated lines near 375 MeV. 
In PQMVT-I model,the $f_0$ becomes $\sigma_S$ and $\sigma$ becomes $\sigma_{NS}$.
These two lines merge with each other near very high temperature of 720 MeV 
.In the PQMVT-II model also, the $f_0$ becomes $\sigma_S$ and $\sigma$ becomes
$\sigma_{NS}$ but these two lines merge near T= 385 MeV.

\begin{acknowledgements}
Computational support of the computing facility which has been developed by the Nuclear Particle
Physics group of the Physics Department, Allahabad University under the Center of Advanced Studies (CAS) 
funding of UGC, India, is acknowledged.
\end{acknowledgements}

\end{document}